\documentclass[twoside,twocolumn,9pt,table]{article}
\usepackage{extsizes}
\usepackage[super,sort&compress,comma]{natbib} 
\usepackage[version=3]{mhchem}
\usepackage[left=1.5cm, right=1.5cm, top=1.785cm, bottom=2.0cm]{geometry}
\usepackage{balance}
\usepackage{mathptmx}
\usepackage{sectsty}
\usepackage{graphicx} 
\usepackage{lastpage}
\usepackage[format=plain,justification=justified,singlelinecheck=false,font={stretch=1.125,small,sf},labelfont=bf,labelsep=space]{caption}
\usepackage{float}
\usepackage{fancyhdr}
\usepackage{fnpos}
\usepackage[english]{babel}
\addto{\captionsenglish}{%
  
}
\usepackage{array}
\usepackage{droidsans}
\usepackage{charter}
\usepackage[T1]{fontenc}
\usepackage[usenames,dvipsnames]{xcolor}
\usepackage{setspace}
\usepackage[compact]{titlesec}
\usepackage{hyperref}

\usepackage{epstopdf}

\definecolor{cream}{RGB}{222,217,201}
\usepackage{xcolor}
\usepackage{changepage}
\usepackage{fontawesome}
\usepackage{lineno}
\begin{document}

\pagestyle{fancy}
\thispagestyle{plain}
\fancypagestyle{plain}{
}

\newcommand{\val}[1]{\textcolor{black}{#1}} 
\newcommand{\LB}[1]{\textcolor{black}{#1}}
\newcommand{\FG}[1]{\textcolor{black}{#1}}
\newcommand{\CB}[1]{\textcolor{black}{#1}}
\newcommand{\SH}[1]{\textcolor{black}{#1}}


\makeFNbottom
\makeatletter
\renewcommand\LARGE{\@setfontsize\LARGE{15pt}{17}}
\renewcommand\Large{\@setfontsize\Large{12pt}{14}}
\renewcommand\large{\@setfontsize\large{10pt}{12}}
\renewcommand\footnotesize{\@setfontsize\footnotesize{7pt}{10}}
\makeatother

\renewcommand{\thefootnote}{\fnsymbol{footnote}}
\renewcommand\footnoterule{\vspace*{1pt}%
\color{cream}\hrule width 3.5in height 0.4pt \color{black}\vspace*{5pt}} 
\setcounter{secnumdepth}{5}

\makeatletter 
\renewcommand\@biblabel[1]{#1}            
\renewcommand\@makefntext[1]%
{\noindent\makebox[0pt][r]{\@thefnmark\,}#1}
\makeatother 
\renewcommand{\figurename}{\small{Fig.}~}
\sectionfont{\sffamily\Large}
\subsectionfont{\normalsize}
\subsubsectionfont{\bf}
\setstretch{1.125} 
\setlength{\skip\footins}{0.8cm}
\setlength{\footnotesep}{0.25cm}
\setlength{\jot}{10pt}
\titlespacing*{\section}{0pt}{4pt}{4pt}
\titlespacing*{\subsection}{0pt}{15pt}{1pt}
\graphicspath{Pictures}


\makeatletter 
\newlength{\figrulesep} 
\setlength{\figrulesep}{0.5\textfloatsep} 

\newcommand{\topfigrule}{\vspace*{-1pt}%
\noindent{\color{cream}\rule[-\figrulesep]{\columnwidth}{1.5pt}} }

\newcommand{\botfigrule}{\vspace*{-2pt}%
\noindent{\color{cream}\rule[\figrulesep]{\columnwidth}{1.5pt}} }

\newcommand{\dblfigrule}{\vspace*{-1pt}%
\noindent{\color{cream}\rule[-\figrulesep]{\textwidth}{1.5pt}} }

\makeatother

\twocolumn[
   \begin{@twocolumnfalse}
\vspace{1em}
\sffamily
\begin{tabular}{m{4.5cm} p{13.5cm} }

\vspace{0.3cm} & \vspace{0.3cm} \\

\bigskip

&

\noindent\LARGE{\textbf{\SH{Discriminating between individual-based models of collective cell motion in a benchmark flow geometry using standardised spatiotemporal patterns }}} 

\bigskip

\noindent\large{Carine Beatrici,\textit{$^{a,b}$} C\'assio Kirch,\textit{$^{a}$}, Silke Henkes, \textit{$^{c}$}, Fran\c{c}ois Graner, \textit{$^{b}$} Leonardo Brunnet \textit{$^{a}$}} \\

& \bigskip \noindent\normalsize{Collectively coordinated cell migration plays a role in tissue embryogenesis, cancer, homeostasis and healing. To study these processes, different cell-based modelling approaches have been developed, ranging from lattice-based cellular automata to lattice-free models that treat cells as point-like particles or extended detailed cell shape contours. In the spirit of what Osborne et al. [PLOS Computational Biology, (2017) \textbf{13}, 1-34]
did for cellular tissue structure simulation models, 
we here compare five simulation models of collective \FG{cell migration}, chosen to be representative in increasing order of included detail. \FG{They are}
Vicsek-Gr\'egoire particles, Szab\'o-like particles, \SH{self-propelled Voronoi model}, cellular Potts model, and multiparticle cells\FG{, where each model includes cell motility}. We examine how these models compare when applied to the same biological problem, and what differences in behaviour are due to different model assumptions and abstractions. For that purpose, we use a benchmark that discriminates between complex material flow models, and that can be experimentally approached using cell cultures: the flow within a channel around a circular obstacle, that is, the geometry Stokes used in his historical 1851 experiment. For each model we explain how to best implement it; vary cell density, attraction force and alignment interaction; draw the resulting maps of velocity, density and deformation fields; and eventually discuss its respective advantages and limitations.
We thus provide a recommendation on how to select a model to answer a given question, and we examine whether models of \FG{motil}e particles and \FG{motil}e cells display similar collective effects.
} \\
\end{tabular}

 \end{@twocolumnfalse} \vspace{0.6cm}
 ]

\renewcommand*\rmdefault{bch}\normalfont\upshape
\rmfamily
\section*{}
\vspace{-1cm}


\footnotetext{\textit{$^{a}$~Instituto de F\'\i sica,
  Universidade Federal do Rio Grande do Sul, Av. Bento Gon\c{c}alves
  9500, C.P. 15051 - 91501-970 Porto Alegre, RS, Brazil }}
\footnotetext{\textit{$^{b}$~Universit\'e Paris Cit\'e, CNRS, Mati\`ere et Syst\`emes Complexes, F-75006 Paris, France. }}
\footnotetext{\textit{$^{c}$~Leiden Institute of Physics, Leiden University, Niels Bohrweg 2, Leiden, NL-2333 CA, The Netherlands. }}
\footnotetext{\ddag~
Simulation codes are available at 
\href{https://github.com/carinebeatrici/Cellular_Stokes_Flow_Simulations}{Stokes Flow Simulations}.}





\section{Introduction}
Collectively coordinated cell migration plays a role in tissue embryogenesis, pattern formation, cancer, homeostasis, regeneration and healing~\cite{Vedula2013,Stock2021}. It  is a \LB{ubiquitous} process involving different morphologies and mechanisms in different cell types and tissue environments~\cite{Friedl2004}. 
Cells grow, move, divide or die, and also change size, shape or neighbours: all these processes contribute together to tissue shape and size changes~\cite{Etournay2015,Guirao2015,Green2022} and generate stresses. Due to the cumulative effects of structural changes at subcellular and cellular levels, the tissue-scale response to these stresses is complex in terms of viscoelasticity~\cite{Tlili2020,Pajic2021}, yielding and jamming~\cite{Bi2016,Hopkins2022}.

Statistical physics and hydrodynamics approaches in active matter studies~\cite{Vicsek1995,Gregoire2003,Ramaswamy2010,Marchetti2013} have 
raised fundamental questions regarding symmetry breaking at the onset of cell migration, either in general~\cite{Huang2005,Stock2021} or in specific cases~\cite{Streichan2011,Weber2012}. Other questions include  motility-induced phase separation and its link to tissue glassiness~\cite{Paoluzzi2022}, and the onset of waves~\cite{Tlili2018} or vortices~\cite{Segerer2015}.

Individual-based numerical models of \FG{motile} cells have been developed in several contexts, each one with its own variants, in two and/or three dimensions.
Some models link the cell scale with collective cell migration~\cite{Albert2016}, and a minority also include the subcellular scale~\cite{Buttenschon2020}, pointing to  cell \FG{motil}ity and polarization as essential ingredients in tissue dynamics.
Other models use the cell center as degree of freedom; in this case a cell is either treated as a point~\cite{Vicsek1995,Szabo2006,Sepulveda2013}, an elastic adhesive circle or sphere~\cite{Hoehme2010,Frascoli2013,rojas2023} or a polygon of a Voronoi tessellation~\cite{Bi2016,Barton2017}. 
Finally, some models describe the cell body in more detail, using the cell contour shape as degree of freedom. This includes descriptions based on vertices of polygons tiling the space~\cite{Tlili2019,Perez2020}, pixels similar to experimental images (cellular Potts model)~\cite{Kafer2006,Kabla2012,Guisoni2018}, several vertices free to move and interacting pairwise~\cite{Teixeira2021}, or a smooth and continuous phase field~\cite{Loewe2020,Ophaus2018}. Finally, some  of these models are lattice-based while others are lattice-free. 




\FG{An exhaustive list of commonly used models in literature is out of scope of this paper. In fact, ``no one review paper can do justice to the entire field'', as claimed by a recent review~\cite{Buttenschon2020}. In order to keep the computational cost reasonable, we have to perform choices. 
Our primary objective is to explore models covering the spectrum from entirely particle-like models to entirely cell-like ones.  We  thus leave the exploration of other important models, in particular the family of phase field ones, for future work.}

Each of \FG{the models we consider here} derives from existing 
non-\FG{motil}e cell simulation models reproducing tissue structure and simple dynamics. Several reviews exist, including those of Fletcher and coworkers~\cite{Osborne2017,Fletcher2017,Fletcher2021}. The two-dimensional version of five models (cellular automaton, cellular Potts model, overlapping spheres, Voronoi tesselation, vertex model) have been compared, using a common computational framework and four case studies~\cite{Osborne2017};  the influence of cell proliferation, adhesion, death, differentiation and signaling range have been studied in detail, and practical conclusions are drawn regarding the choice of a model to address a given question. 

\begin{figure*}[h]
\begin{center}
\includegraphics[height=0.40\textwidth]{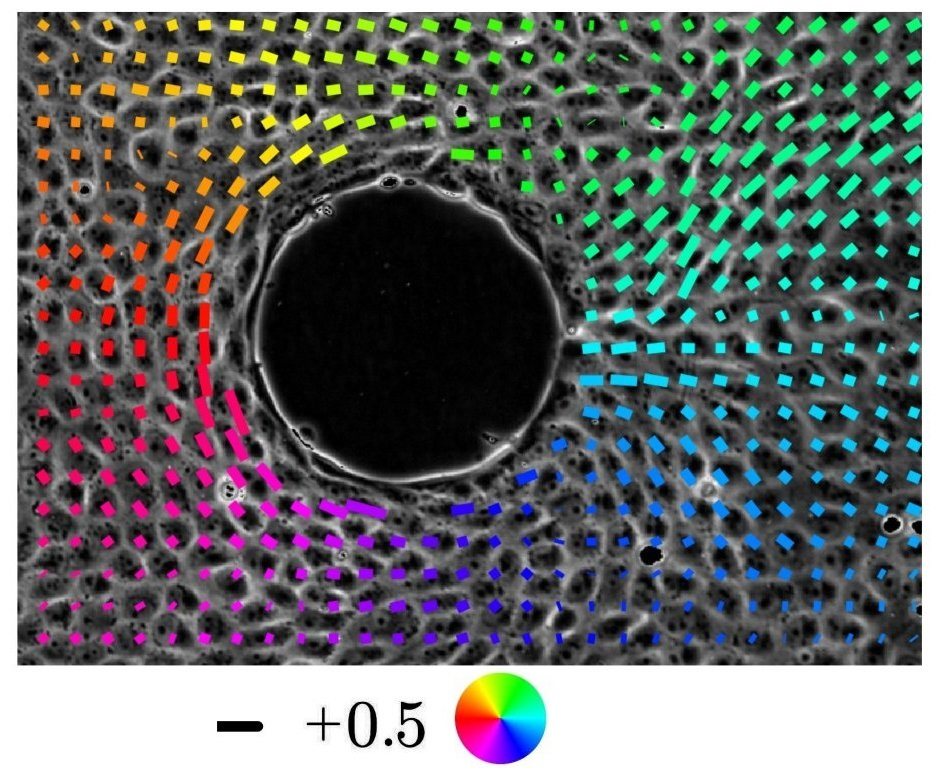}
\includegraphics[height=0.395\textwidth]{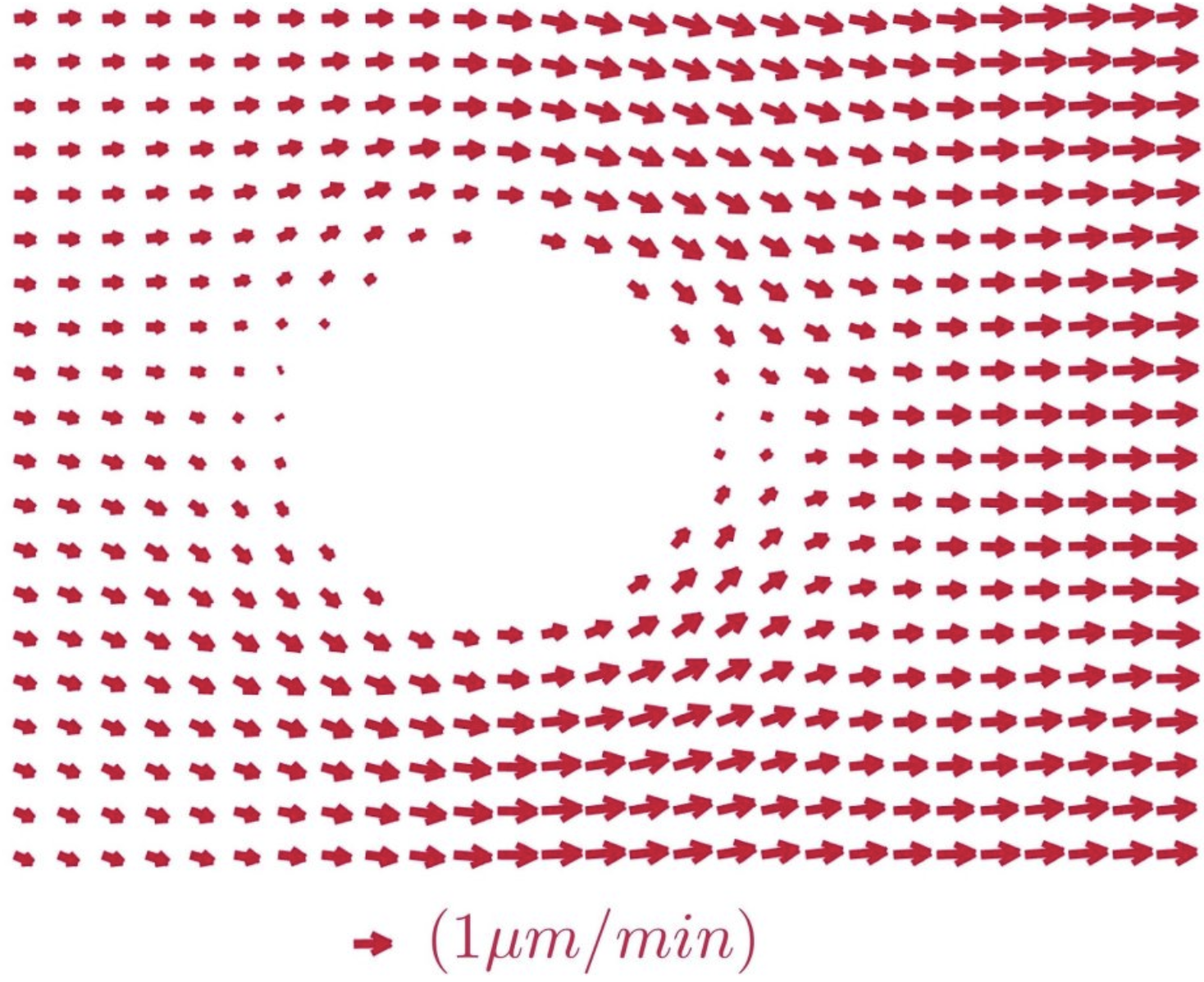}
\caption{Experiment. A  MDCK epithelial cell monolayer after 20 hours of migration in Stokes  geometry, from left to right. On a flat substrate, a channel is drawn as a region where cells can adhere and crawl, while the channel walls and a circular obstacle are \SH{unfavorable} for cell adhesion. 
\FG{(a) Deformation field measured for a phase contrast snapshot. The deformation tensor deviator  is diagonalized and each bar represents its main axis of extension. 
The color codes for the angular position of each point, in polar coordinates originating at the obstacle center. (b) Corresponding velocity field  averaged over 8 h.
Scales are indicated below each panel. Maps are zoomed around the obstacle; actual strip length 4 mm, strip width 1 mm, obstacle diameter 0.2 mm, pixel size 0.65~$\mu$m.
Reproduced with permission from Ref.~\cite{Tlili2020}, which did not publish the corresponding density field.}
}
\label{fig:exp_snap}
\end{center}
\end{figure*}

\LB{
In the context of cell tissues, the flow around a circular obstacle in a two-dimensional channel \FG{can play} a significant role. This geometry favors shear and viscous flow, which is essential for understanding the heterogeneous deformation, deformation rate, and rearrangement rate of cells. These non-zero velocity gradients, resulting from the heterogeneity of cell velocity orientations, are critical for gathering the discriminant information \FG{regarding both amplitude and direction (so-called ``tensor field" information)} necessary for benchmarking cell migration models. Moreover, the flow around a circular obstacle was used in Stokes' historical 1851 experiment~\cite{Stokes1851} and is similar to the motion of an intruder within a cellular material~\cite{Hopkins2022}. For non-\FG{motil}e cellular materials like soap froth, this geometry \FG{has been particularly efficient in} differentiating and testing different models' predictions~\cite{Cheddadi2011}.  The corresponding experiment with cells is feasible  and has been carried out several times~\cite{Kim2013,Tlili2020,Durande2020}\FG{, as illustrated in Fig.~\ref{fig:exp_snap}. Quantitative comparison between experiments and models is beyond the scope of this study and will be the subject of future work.} 
}

\begin{figure*}[h]
    \centering
    \includegraphics[width=0.8\textwidth]{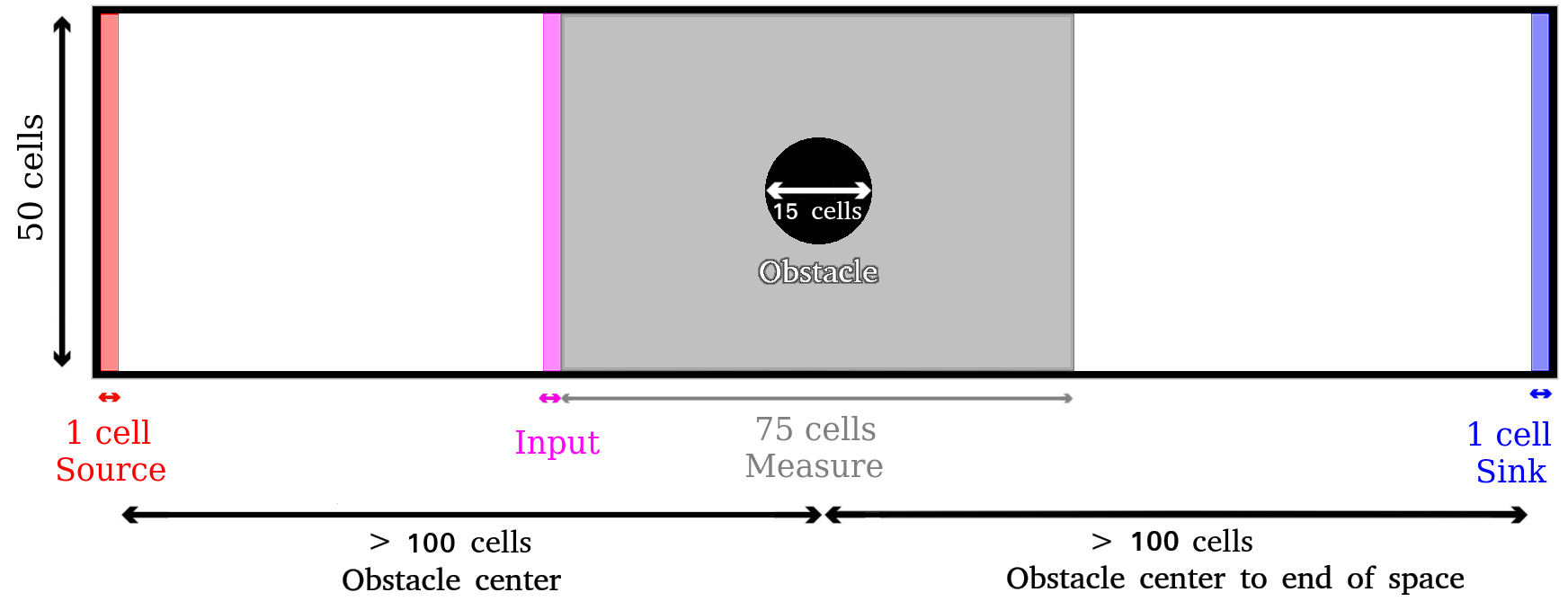}
    \caption{\FG{Simulation set-up, definitions and dimensions (expressed in cell size units).}}
    \label{fig:stokes_geometry}
\end{figure*}
Here, we build on these advances in simulation models and of benchmarking. In the spirit of Ref.~\cite{Osborne2017}, we run comparable collective movement simulations in the Stokes geometry for five \FG{motil}e cell simulation models, chosen as representative of the  
progression from the simplest to the most detailed. The first model is derived from the now classical Vicsek and Gr\'egoire particle models~\cite{Vicsek1995,Gregoire2003}. The second one, the Szab\'o-like particle model~\cite{Szabo2006}, is similar, but dominated by cell velocity self-persistence instead of direct neighbor alignment. The third one, \SH{the self-propelled Voronoi model, is} based on particles
associated with a Voronoi tesselation~\cite{Bi2016,Barton2017}, is chosen because \LB{it is} an intermediate between cell center and cell contour based models. The fourth one, derived from the cellular Potts model~\cite{Kafer2006,Kabla2012}, uses pixels and thus an experimental image can be directly compared with simulations (or even injected as the initial image of a simulation~\cite{Bardet2013}). The last one, which uses multiparticle cells~\cite{Teixeira2021}, can handle highly deformed cells and the dissipation associated with cell shape changes. For all five models, and especially for the fifth, we have introduced new details with respect to the literature.

Our motivation is twofold. First, we want to understand how each model behaves, depending on its ingredients and underlying assumptions, and examine the common points and differences between models. In particular, models based on cell centers versus on cell contours display common properties (e.g. soft elastic particles versus self-propelled Voronoi~\cite{Henkes2020}) but it is unclear to which extent. Second, we want to examine the respective advantages and limitations of each model: for each given scientific question we want to provide the reader with a guide to help choosing the most adequate model, the best implementation method and the range of parameter values. For that purpose, we vary input parameters such as cell density, attraction force and alignment interaction; as outputs we draw the resulting maps of velocity, density and deformation fields. 

This paper is organized as follows. We first describe the common simulation set-up, the choice of parameters and measurements and present the formulation of the five models and their implementation. We then present the results for each model, that is the input parameter range and the output measurement maps. We compare and discuss these results, along with a guide for the reader (Table~\ref{tab:model_properties_comparison}), and conclude.

\subsection{Simulation set-up}
\label{setup}

\CB{
Our benchmark is a standard simulation set-up common to the five models. Cells flow within a channel around a circular obstacle (Stokes geometry)~\cite{Cheddadi2011}. 
To keep cells migrating and to emulate a steady-state-like
regime, we constantly create new cells in the source region on the left side of the channel, in red on Fig.~\ref{fig:stokes_geometry}, and drop cells at the same rate in the sink 
region on the right side of the channel, in blue on Fig.~\ref{fig:stokes_geometry} (with a few variations for the Voronoi model). 
}

\CB{
The cell diameter at equilibrium may depend on several model parameter values such as
the force between neighboring cells, or the cell
creation rate. In order to compare simulations, we use the cell equilibrium diameter as the unit length.
In these units, the channel is 50~cells wide and the obstacle diameter is 15~cells while source and sink regions are only 1 cell long.
}

\CB{
The simulations produce snapshots over which we make two very different sets of measurements, which we call ``input measurements'' and ``output measurements''. Output measurements are our results, and are plotted as maps over the whole output measurement region, which
is 75~cells long, and centers on the obstacle.
Conversely, input measurements are used to monitor the simulation at the entrance of the output measurement region, and ensure the comparison between different models is performed in similar conditions. 
The input measurement region is 1 cell long and the spatial average is performed over the channel width.
}

\CB{
Close to the source region, the creation process  
 frequently produces transient artifacts  which can vary from model to model, which motivates us to leave a model-dependent transition region between the source and the input measurement region.
We set the obstacle
center at least 100~cell diameters from the 
cell source region and we use the same distance from the
obstacle center to  the sink region.
After a transient period to allow for the steady-state-like regime to establish itself, with a time scale determined by the typical cell velocity divided by the obstacle size, measurements are averaged in time over the simulation duration.
}

\subsection{Acceptable parameter values}


\LB{Each model has its own restrictions when it comes to acceptable parameter values. Our objective is to determine these values and identify the specific regions for each model where realistic cell flow can potentially occur.
}

\begin{figure*}[h]
\begin{center}
\includegraphics[width=0.8\textwidth]{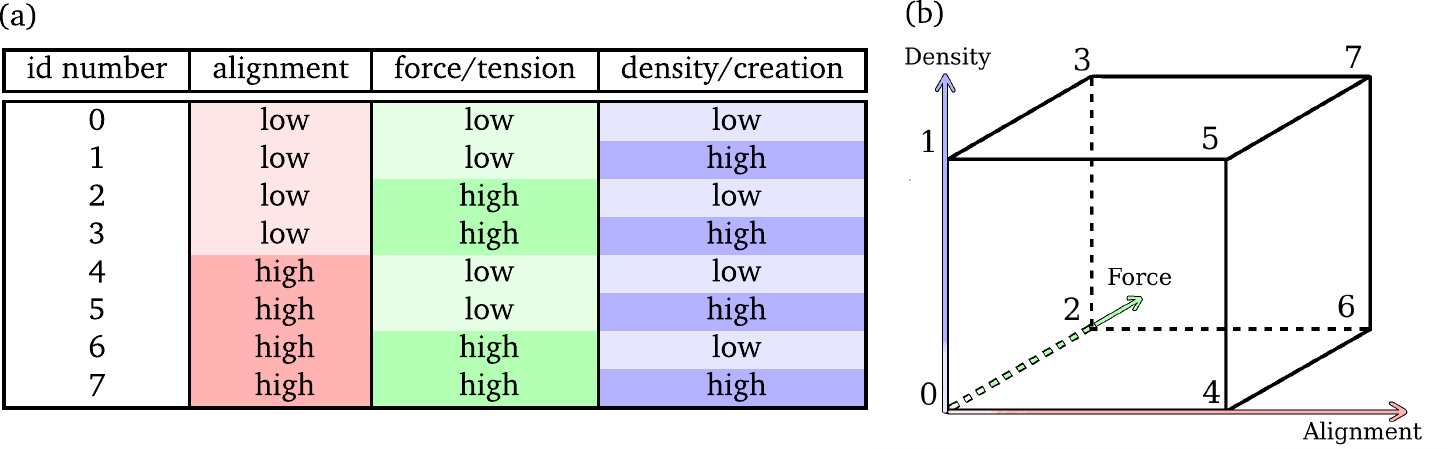}
\caption{Representations \SH{of parameter space}. (a) \FG{The id numbers 0, 1, 2, ... 7, corresponding to the corners of a three-dimensional cube, identify the limits of parameters for the simulations and
    resulting maps.} (b) Cube \FG{visualising} extreme value labels, as defined in (a). 
}
\label{fig:limits-and-cube}
\end{center}
\end{figure*}

We are interested in three main model parameters:  alignment (that affects
the collective migration), force/tension between neighbor cells
(that affect the tissue liquid or solid behaviour) and cell creation rate
(that affects the density). In Fig.~\ref{fig:limits-and-cube},
the eight limit cases are presented
and identified by a number 0, 1, 2, ... 7 which we use throughout this article.

Some parameter limits are simply due to the numerical 
implementation, as the numerical 
solution may not converge, or the
simulation may stop running due to infinite or non numerical values.
Other more striking limitations are the physical and biological ones, like 
unrealistic densities or velocities. 
For example, the particles with Voronoi model can not support
empty spaces; therefore for low densities, instead of creating
empty spaces in the tissue the cells 
would stretch indefinitely.
In many cases, some parameter values may 
generate artifacts in the dynamics and 
the physics is no longer 
correct.

\subsection{Input measurements}
\label{input_measurements}

\subsubsection{Implementation of input measurements.}

A natural approach to compare simulations from different 
models would consist in standardizing
the set of parameters from the different models in order to 
construct a common set of dimensionless numbers based on the model parameter values~\cite{Belmonte2008}.
However, here, this approach is unfeasible: 
In fact, model ingredients are very diverse, especially since cell centers and cell contours are qualitatively different degrees of freedom. Even the number of model parameters varies a lot, from the parsimonious Vicsek-Gr\'egoire model to the detailed Potts model, so that the number of relevant dimensionless parameters would be difficult to decide. 

We have therefore chosen an alternative route: we define a standardized set of dimensionless \emph{input measurements}.
This has the following advantages:
First, we can draw a common phase diagram, with identical axes corresponding to input measurements; we can  then position each simulation on these axes, and thus on the same phase diagram. 
Second, if in the future a reader wants to compare the current five simulation models with a new one, it will not be necessary to perform any theoretical analysis; it will be sufficient to measure the input quantities as we do here. 
Third, it will determine
which models can or cannot be compared; if the input measurements do not present any
intersection range, the models are 
too different to be comparable. 
Fourth, the input measurements are physical quantities and are in direct correspondence with the output measurements that we are interested in. In contrast, some model ingredients have no intuitive physical interpretation, or are not in correspondence with the output measurements.
Fifth, the same approach will in principle be applicable to experiments too; in fact, the input measurements  are 
accessible from experiments, as opposed to the dimensionless numbers based on the underlying parameter values. 

Here, given our interests in the cellular and tissue aspects, we choose as input measurements three cell-scale characteristics:
First, the {\it alignment} of a cell velocity with its neighbours velocity, which quantifies local order or disorder in the velocity field. Second, the {\it liquid or solid behaviour}, based on each cell center's  local displacements relative to its neighbors. 
  Third, the {\it relative density} that characterizes the confluence and compression of the monolayer, or its absence.

\LB{Additional \SH{measures}, such as \SH{tissue} softness and viscosity \cite{Liu2014}, have been shown to provide valuable insights into cell behavior \cite{Liu_2021, Lv_2020}. However, due to the variety of particle-based models (Vicsek, Szabó) and extended models (Voronoi, Potts, Multiparticle) utilized in our study, these \SH{measures} cannot be readily applied to all models. Furthermore, as commented above, each model that we have used has additional parameters to vary beyond the three we have chosen, and these cannot be easily mapped to one another within each model. Therefore, we have selected the most common values reported in the literature for these fixed parameters.}
 Below, we examine each of the three chosen quantities in greater detail, and show how to measure them in practice. 
 
For each model we determine the set of 
ingredients that can contribute to set 
these particular
tissue characteristics;  these ingredients are model-dependent. 
For instance, in some models the alignment is explicitly prescribed, while in others it is only an indirect consequence of ingredient choices. The cell behaviour can become more solid-like due to a large interaction force between cell centers, or to a large tension of cell-cell junctions. The density can directly or indirectly depend on several ingredients, for instance it increases with the cell creation rate (when it exists) and decreases with the free cell velocity.  

We run simulations with several values of the model parameters to delimit the accessible range of input measurements. 
The phase diagram is three dimensional so that there are eight combinations of limit cases which we explore (Fig.~\ref{fig:limits-and-cube}).
Note that in principle, there can be several combinations of model parameter values that result in the same limit case.
Exploring these combinations of parameter values is  beyond the scope of this work. 
 Here, we choose to change as few parameter values as
possible at a time, ideally one. 

\subsubsection{Choice of dimensionless input measurements.}

 
To measure the degree of alignment of \FG{motil}e cell movements, we use the parameter originally proposed by Vicsek \textit{et al.}~\cite{Vicsek1995},  the velocity order parameter: 
\begin{equation}
    \phi = \frac{1}{N} \displaystyle \sum_{i \in N} \frac{\vec{v}_i}{|v_i|} 
    \label{phi}
\end{equation}
where $N$ is the number of cells and $\vec{v_i}$ is the velocity of cell $i$. 
If each cell movement direction is uncorrelated 
with the surrounding ones, $\phi = 0$, cells form a non-collective  flow.
Conversely, if cells are all moving 
in the same direction, $\phi = 1$, they form a collective flow.


In order to evaluate the level of solidity or liquidity in the tissue, we could employ a measure commonly used in the soft matter field: the mean-square displacement ($MSD$). This measure quantifies the distance that a particle travels in time $t$, and is averaged spatiotemporally, i.e., over both space and time, denoted as $\langle (\mathbf{r}(t_0 + t)-\mathbf{r}(t_0))^2 \rangle_{t_0,\text{space}}$. When $MSD$ reaches a value around $\sigma^2$, where $\sigma$ represents the typical size of cells, it indicates the occurrence of a glass-to-liquid transition. Nevertheless, this measure is only an indirect indicator of rearrangements, and it is sensitive to  spatial irregularities and the method of overall flow subtraction.

We thus choose here to use the more robust parameter
proposed by Gr\'egoire \textit{et al.}~\cite{Gregoire2003}:
\begin{equation}
    \Delta = 1 -  \frac{1}{n_i} \displaystyle \sum_{i \sim j} \left( 1 -  \frac{r^2_{ij}(t)}{r^2_{ij}(t+T)} \right) 
    \label{delta}
\end{equation}
where $r_{ij}(t)$ is the distance between 
centers of cells $i$ and $j$ at time $t$, while $r_{ij}(t+T)$ 
is their distance after time interval $T$. This sum is
normalized by the number of particles $n_i$.  
By that definition $\Delta$ is close to one
when  a cell's motion is only fluctuating 
locally, keeping most of its neighborhood: this is solid-like behaviour.
Conversely, $\Delta$ is close to zero
when a cell frequently exchanges most of its neighborhood: this is liquid-like behaviour.
The value of $\Delta$ of course depends on the choice of $T$, and this point is even more sensitive
for an out-of-equilibrium tissue like the one we consider here. 
To choose $T$, we use an adaptive method: 
We first run the transient simulation
time steps, and list the cells 
inside the input measurement region.
We then track them while they flow 
over one obstacle radius and calculate  $\Delta$ during this time interval $T$.  Using the measurement over an interval of \LB{one} obstacle radius just beyond the region of input measurements allows us to define whether the cells exchange their neighborhood along a spatially well-defined region, sufficiently far from the source and the obstacle, and independent of the velocity associated with the flow.

In order to convert the density into a non-dimensional form, we define
\begin{equation}
\delta\rho = \left\langle \frac{\rho}{\rho_{eq}} \right\rangle - 1,
\label{rho}
\end{equation}
where $\rho$ denotes the number of cells per unit area, and $\rho{eq}$ represents its equilibrium value under model-specific conditions, i.e., in the absence of external forces and stresses.
According to this definition, $\delta\rho$ vanishes when, on average, the cells
are at equilibrium 
density. \LB{It becomes positive when the cells are compressed, and negative when the cells are stretched or create gaps.}

\subsection{Output measurements}

Output measurements are performed over 466 boxes disposed in a  28 $\times$ 18 rectangular grid (minus 38 grid elements corresponding to the obstacle). 
We measure and represent the following three quantities.

The normalized density $\delta\rho = 0$ is the same as the one used as an input measurement  (Eq. \ref{rho}). It is a scalar quantity and is represented by a color. Blue represents negative values of $\delta\rho$, i.e. density lower than the 
equilibrium; white represents $\delta\rho = 0$, i.e. density at equilibrium; and red represents positive values of $\delta\rho$, i.e.  density higher than in equilibrium. 

For each snapshot, we measure each cell velocity during the time interval  immediately following the snapshot. 
The velocity, averaged over all cells in the box, is a vector represented as an arrow  which we place 
in the middle of the box, while a yellow unit scale arrow is shown in the middle of the obstacle. 
 In the snapshot, we color each particle by its direction of movement according to the angular color map shown in Fig.~\ref{fig:1-particle-simul}.

The deformation is the anisotropy of the coarse-grained cell shape deformation (not to be confused with the coarse-grained average of the cell shape deformation anisotropy).
It is measured by averaging links between cells using the inter-cellular texture matrix as defined in reference~\cite{Graner2008}. We divide the system into boxes of four cell diameters in size and perform a time average of the textures over typically 50 snapshots, during which cells have moved at least 30 cell diameters. The average texture is diagonalized yielding two eigenvalues, $L_{max}^2$ and $L_{min}^2$. From these we calculate the cell deformation deviator amplitude, $\frac{1}{2}\ln{\frac{L_{max}}{L_{min}}}$, and the cell deformation deviator orientation, which  is the angle of the larger eigenvalue direction relative to the $x$-axis~\cite{Durande2019}. We represent the deviator as a bar,  with a length corresponding to the magnitude of the deformation anisotropy and with an angle corresponding to its major axis. 
To indicate scale, the red line in the middle of the obstacle represents a deformation of $\ln 2$, corresponding to cells whose length is twice their width. 

\section{Materials and Methods: Simulation Models}
\label{models}

In this section we present the simulation 
models covering their principle, their implementation, and their parameters.
Ingredients include \FG{motil}ity, alignment, 
polarization, interaction (force between cell centers, or cell-cell junction tension),
area, perimeter, density, cell creation and cell destruction.
We emphasize that all models are in their \FG{motil}e version.

For each model, we determine three  model parameters that affect the three input measurements 
alignment (Eq. \ref{phi}), rigidity (Eq. \ref{delta}) and density relative to the equilibrium density (Eq. \ref{rho}). 
We vary these three model parameters (keeping the others fixed) and determine the range of their values which lead to
low and high levels of these input measurements. We also briefly discuss the effects on running simulations outside of this parameter 
range. 


\subsection{Vicsek model}

The Vicsek model~\cite{Vicsek1995} describes each cell $i$ as a single \FG{motil}e particle. 
For each time step, the particle position evolution is given by
\begin{equation}
\vec{x}_i(t+\Delta t)=\vec{x}_i(t)+\vec{v}_i(t)\Delta t.
\end{equation}
Here, the time interval is fixed as 1 and the time scale
is determined by the velocity module, chosen as $v_0 = |\vec{v}_i| = 0.05$.
Each particle has a  speed of fixed modulus, so it always 
moves regardless of the external forces and all particles are identical.

The sole degree of freedom is the velocity direction, which evolves according to~\cite{Gregoire2003}:
\begin{equation}
\theta_i (t+\Delta t) = \arg \left[\displaystyle\sum_{j\sim \left\langle i \right\rangle }
\alpha \frac{\vec{v}_j(t)}{v_1}
+ \displaystyle \sum_{j\sim \left\langle i \right\rangle } \beta \vec{f}_{i,j}(t) + \eta
\vec{u}_i (t)\right]
\end{equation}

The first term is the alignment with neighbors, here an explicit model ingredient.
These neighbors are defined according to a metric (i.e. distance-based, as opposed to topology-based) criterion where 
$j$ is neighbour to $i$ if their distance is smaller than a distance $r_{max} = 1$.
The collective migration behaviour is then tuned by the coupling parameter $\alpha$.

The second term is the pairwise, radial force between neighboring particles,
tuned by the $\beta$ coupling parameter:
\begin{eqnarray}
f_{i,j} = \left\{ \begin{array}{cl} 0 & \;\;\;\;\;\;\;\; r_{ij} \ge r_{\text{max}} \\ 1 -
\frac{r_{ij}}{r_{\text{eq}}} & \;\;\;\;\;\;\;\; r_{c} < r_{ij} < r_{\text{max}} \\ + \infty & \;\;\;\;\;\;\;\;
r_{ij} \le r_{c} \end{array} \right.
\label{eq:vicsek_force}
\end{eqnarray}
Particles have a hard-core repulsion ($f_{c}=1000$) of  below a radius $r_{c} = 0.18$.
Between $r_{c}$ and  $r_{max}$ the force is harmonic
and the
equilibrium force distance is $r_{eq} = 0.8$; 
$r_{eq}/2$ is used as the 
size unit. This equilibrium distance we define as the equilibrium density for $\rho_{eq} = 1 /(\pi (r_{eq}/2)^2)$.
The last term is the vector noise where $\vec{u}_i (t)$ is a random unitary vector, and where we keep the amplitude $\eta$ fixed as one.

The system dimensions in simulation units are: 
channel length 101, width 25, obstacle center position (50, 12.5), obstacle radius 3.75, 
source region from $x = 0$ to 1, sink region at  $x=100$.
The time scale 
is given by the particle speed and time interval;
we keep $v_0 \Delta t < 0.1 \; r_{C}$ to prevent a particle from jumping over another one.

\begin{figure*}[h]
\begin{center}
\includegraphics[width=\textwidth]{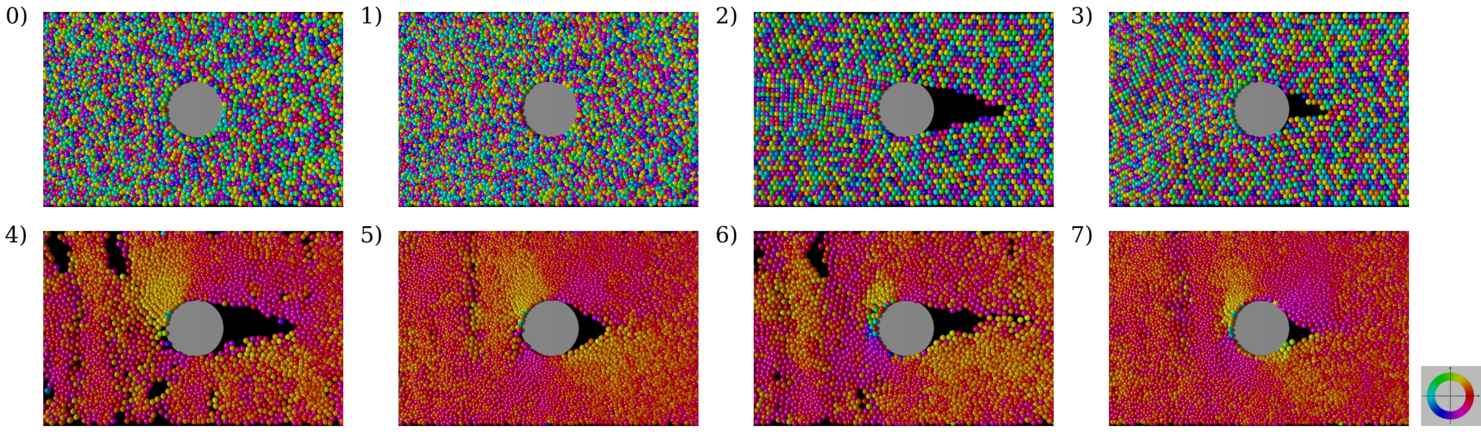}
\caption{Snapshots for the \FG{motil}e cell simulation in the Stokes geometry in the
Vicsek model's eight limit cases where the panel labels correspond to Fig.~\ref{fig:limits-and-cube}.
The values for the parameters used in this model are specified in Table~\ref{tab:param-boid}.
Images with even numbers present systems with density close to confluence, 
while the odd ones are constructed with higher densities. 
The top row presents the low alignment cases while the bottom
one presents the high alignment ones. The four images on the left correspond to low attraction forces (liquid-like), while the four images on the right correspond to high attraction forces 
(solid-like).
The images are restricted to an area around the obstacle; particle source and sink regions are not depicted. 
The color of each particle is related to the direction of its movement\FG{, see the orientational color map at the bottom right.
Objects migrating in the flow direction, along the positive $x$-axis towards the right, are displayed in red. Objects moving upwards, along the positive $y$-axis, are represented in yellowish-green. Objects
moving backwards, towards the left, are cyan. Objects moving downwards are a blueish-purple color. } \SH{A black background color corresponds to holes in the tissue.}
}
\label{fig:1-particle-simul}
\end{center}
\end{figure*}
\begin{table}[h]
    \centering
    \begin{tabular}{|c|c|c|c|}
            \hline 
            Parameter & Level & Value \\
            \hline
            \hline
            Alignment ($\alpha$) & low & 0.0  \\
            Alignment ($\alpha$) & high & 0.5  \\
            Force ($\beta$) & low  & 2.0 \\   
            Force ($\beta$) & high & 5.0 \\
            Creation ($rate$) & low & 0.007 to 3.0 \\
            Creation ($rate$) & high  & 0.0105 to 5.0 \\
            \hline
    \end{tabular}
    \caption{Limit values for the parameters varied in the Vicsek  model.
    The particle creation rate needs to be carefully adjusted in order to keep the flow as confluent 
    and steady as possible, and this adjustment strongly depends on the alignment degree.  
    The lowest creation rate to keep confluent flow is $0.007$ for disordered cells, 
    while it is $3.0$ to keep a highly aligned confluent flow. 
    In order to produce a high density flow, we increase creation rate by approximately $50\%$, 
    which leads to the high creation rate of $0.0105$ for disordered cells and $5$ for ordered cells.
    }
    \label{tab:param-boid}
\end{table}

\begin{table}[h]
    \centering
    \begin{tabular}{|c|c|c|c|}
        \hline 
        id & $\phi$ & $\Delta$ &  $\delta\rho$\\
        \hline
        \hline 
        0 & \cellcolor[rgb]{1.0,0.9,0.9} 0.124 &
        \cellcolor[rgb]{0.9,1.0,0.9} 0.025 &
        \cellcolor[rgb]{0.9,0.9,1.0} 0.711 \\
        
        1 & \cellcolor[rgb]{1.0,0.9,0.9} 0.122 &
        \cellcolor[rgb]{0.9,1.0,0.9} 0.018 &
        \cellcolor[rgb]{0.7,0.7,1.0} 1.022 \\
        
        2 & \cellcolor[rgb]{1.0,0.9,0.9} 0.158 &
        \cellcolor[rgb]{0.7,1.0,0.7} 0.686 &
        \cellcolor[rgb]{0.9,0.9,1.0} 0.062 \\
        
        3 & \cellcolor[rgb]{1.0,0.9,0.9} 0.154 &
        \cellcolor[rgb]{0.7,1.0,0.7} 0.623 &
        \cellcolor[rgb]{0.7,0.7,1.0} 0.381 \\
        
        4 & \cellcolor[rgb]{1.0,0.7,0.7} 0.967 &
        \cellcolor[rgb]{0.9,1.0,0.9} 0.824 &
        \cellcolor[rgb]{0.9,0.9,1.0} 0.174 \\
        
        5 & \cellcolor[rgb]{1.0,0.7,0.7} 0.990 &
        \cellcolor[rgb]{0.9,1.0,0.9} 0.892 &
        \cellcolor[rgb]{0.7,0.7,1.0} 0.997 \\
        
        6 & \cellcolor[rgb]{1.0,0.7,0.7} 0.963 &
        \cellcolor[rgb]{0.7,1.0,0.7} 0.908 &
        \cellcolor[rgb]{0.9,0.9,1.0} 0.169 \\
        
        7 & \cellcolor[rgb]{1.0,0.7,0.7} 0.988 &
        \cellcolor[rgb]{0.7,1.0,0.7} 0.933 &
        \cellcolor[rgb]{0.7,0.7,1.0} 1.014 \\
        
        \hline
    \end{tabular}
    \caption{Input measurements for the Vicsek model. The values of the three input measurements, alignment $\phi$, liquid-solid behaviour $\Delta$ and  normalized density $\delta\rho$, are indicated for the Vicsek model simulations with different values of the three model parameters.
    A lighter color means
    the simulation was performed with a lower level of 
    the parameter related to that measure, a darker color
    means a higher level of that parameter, see Table~\ref{tab:param-boid}. 
    For example,
    the line with id = 3 is the result of a simulation
    with low alignment, high force and high creation rate. }
    \label{tab:boids-cube-result}
\end{table}

Fig.~\ref{fig:1-particle-simul} shows simulation snapshots in the limit cases.
The three model parameters directly affect the input measurements, as shown in Table~\ref{tab:boids-cube-result}. 
First, a low value of the alignment $\alpha$ prevents any collective behaviour (see top row of Fig.~\ref{fig:1-particle-simul}), i.e the $\phi$ value is low. When the alignment $\alpha$ value is high, the collective phase is well simulated, and the $\phi$ value is high.
Note that at high alignment, particles barely separate and $\Delta$ is high, as in a solid phase.
Second, the force $\beta$ determines the liquid versus solid behaviour, $\Delta = 0$ to 1; note the nearly crystalline structure in images 2 and 3 of Fig.~\ref{fig:1-particle-simul}. 
  Finally, the density  $\delta\rho$ increases with the cell creation rate. A low creation rate keeps the density around confluence while a high one keeps the cells under pressure.
 This creation rate needs to be carefully adjusted in order to keep the flow as steady as possible (Table~\ref{tab:param-boid}). 
Note the frequent formation of voids at different parameter values.
Overall, the Vicsek model is robust to parameter 
variation and  artifacts are easy to avoid.

\subsection{Szab\'o model}

The Szab\'o model~\cite{Szabo2006} is also based on \FG{motil}e particles, but is defined as a set of continuous differential
equations, and with no explicit neighbor alignment term.
Each  cell has a polarity direction, which determines self-persistence of the velocity. 
This polarity changes with collisions and with an angular noise. Any collective
behavior in this model is a result of this self-persistence~\cite{henkes2011,baconnier2021}. 

The $i$-th particle polarity $\hat{n}_i$ is a unitary vector with
direction $\theta_i$. 
This angle tends to relax to the direction
of the particle displacement $\vec{v}_i = d\vec{r}_i/dt$ 
in a characteristic time $\tau$:
\begin{equation}
\frac{d\theta_i(t)}{dt} = \frac{1}{\tau}\arcsin\left[\left(\hat{n}_i \times \frac{\vec{v}_i}{|\vec{v}_i|}\right)\cdot \hat{e}_z \right] + \xi_i
\end{equation}
The angular noise $\xi_i$ follows a Gaussian distribution 
with zero mean $\left\langle \xi(t) \right\rangle  = 0$ and auto-correlation 
$\left\langle \xi(t)\xi(t') \right\rangle =\frac{\eta^2}{12} \delta(t,t')$ where $\delta$ is
the Dirac delta function.
There is no direct noise on the displacement, and the angular noise only changes the polarization direction;
$\hat{e}_z$
is the unit vector orthogonal to the plane of motion.

The velocity evolution of the $i$-th particle is given by
\begin{equation}
    \frac{d\vec{r_i}(t)}{dt} = v_0 \hat{n}_i(t) + \mu \sum_{j=1}^{N} \vec{f}(|\vec{r}_{ij}|).
\end{equation}

Without any external influences, the particle will move 
in the polarity direction with its free
velocity $v_0$. 
The interaction with particles or obstacles follows an overdamped Langevin dynamics, where the mobility (or inverse friction) $\mu$ controls the amplitude of the velocity response to forces. Additionally, if force and polarity vectors are aligned, particle velocity increases while it slows down in the converse case. This type of non-reciprocal interaction is responsible for the global flocking state in the system.

The force between two particles $i$ and $j$ is radial, i.e. it only 
depends on their distance $r_{ij} = |\vec{r}_{ij}|$:
\begin{eqnarray}
f(r_{ij}) = \left\{ \begin{array}{cl} 
F_{\text{rep}}\frac{r_{ij}-r_{\text{eq}}}{r_{\text{eq}}} & \;\;\;\;\;\;\;\;  r_{ij} < r_{\text{eq}} \\ 
F_{\text{adh}}\frac{r_{ij}-r_{\text{eq}}}{r_{\text{max}}-r_{\text{eq}}} & \;\;\;\;\;\;\;\;  r_{\text{eq}} \le r_{ij} < r_{\text{max}} \\
 0 & \;\;\;\;\;\;\;\;  r_{ij} \ge r_{\text{max}}.
\end{array} \right.
\end{eqnarray}

At short distance the particles repel each other with a harmonic repulsion with stiffness parameter $F_{\text{rep}}$. 
If the particles are more distant than the equilibrium distance
$r_{\text{eq}}$ they adhere with adhesion parameter $F_{\text{adh}}$, and finally, if the particles are more distant than $r_{\text{max}}$ they do 
not interact. 

The 
fixed parameters are: interaction coupling $\mu = 1.0$, repulsion 
between particles $F_{\text{rep}} = 30.0$, particle free velocity $v_0 = 0.05$, noise amplitude $\eta = 1.0$. The interaction of 
the obstacle with the particles is defined as a central 
repulsive force with stiffness constant equal to $100.0$.
The maximum 
interaction and alignment distance is $r_{\text{max}} = 1.0$, 
and the equilibrium force distance is $r_{\text{eq}} = 0.666$. 
To
avoid crystallization as an artifact of this model we introduce in the equilibrium distance $r_{\text{eq}}$ a polydispersity $0.1$.
 
The system dimensions in simulation units are 
channel length 101, width 25, obstacle center position (50, 12.5), obstacle radius 3.75, 
source region from $x = 0$ to 1, and sink region at  $x=100$.
The time
interval used for numerical integration is $\Delta t=0.005$, chosen for numerical stability and also such that $\Delta t \ll \tau$.

\begin{table}[h]
    \centering
    \begin{tabular}{|c|c|c|c|}
            \hline 
            Parameter & Level & Value \\
            \hline
            \hline
            Alignment ($\tau$) & low & 100.0  \\
            Alignment ($\tau$) & high & 0.1  \\
            Force ($F_{\text{adh}}$) & low  & 1.0 \\   
            Force ($F_{\text{adh}}$) & high & 3.0 \\
            Creation ($div$) & low & 0.01 to 0.35 \\
            Creation ($div$) & high  & 0.015 to 0.525 \\
            \hline
    \end{tabular}
    \caption{Limit values for the parameters varied in the Szab\'o  model.
    The creation rate needs to be carefully adjusted in order to keep the flow as confluent and steady as possible, and
    this adjustment strongly depends on the degree of alignment:  The lowest creation rate to keep confluent flow
    is $0.01$ for disordered cells, while it is $0.35$ to keep a highly aligned confluent flow. 
    In order to produce a high density flow, we increase creation rate by approximately $50\%$, 
    which leads to the high creation rate of $0.015$ for disordered cells and $0.525$ for ordered cells.
    }
    \label{tab:param-szaboid}
\end{table}

\begin{figure*}[h]
\begin{center}
\includegraphics[width=\textwidth]{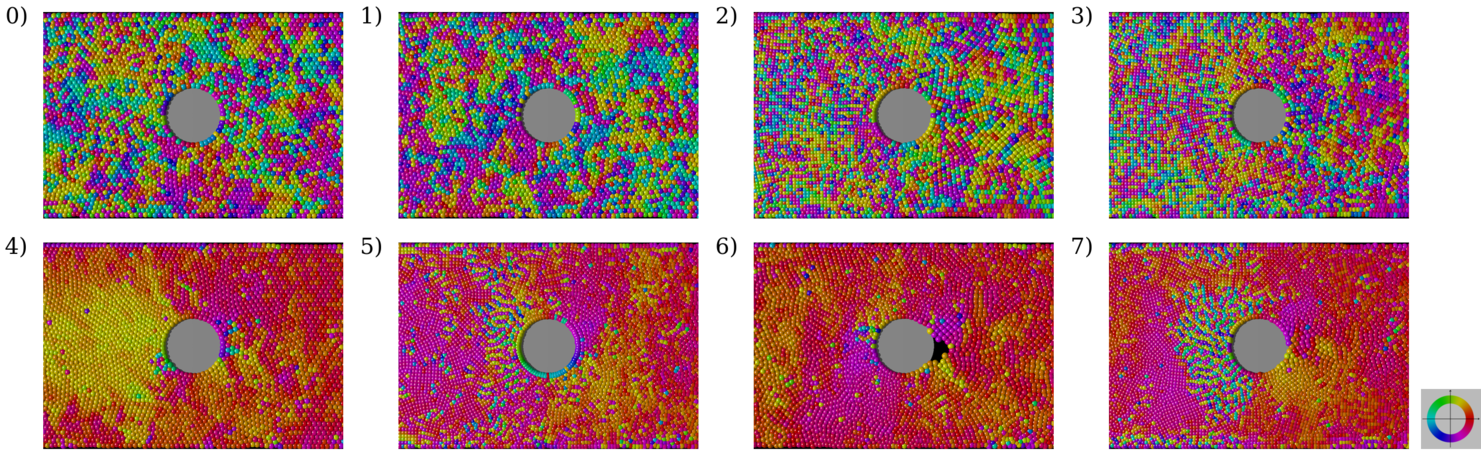}
\caption{Snapshots for the Szab\'o's eight limit cases; same caption as 
Fig.~\ref{fig:1-particle-simul}. See Table~\ref{tab:param-szaboid} for parameter values.
}
\label{fig:snap-szaboid}
\end{center}
\end{figure*}

\begin{table}[h]
    \centering
    \begin{tabular}{|c|c|c|c|}
        \hline 
        id & $\phi$ & $\Delta$ &  $\delta\rho$\\
        \hline
        \hline 
        0 & \cellcolor[rgb]{1.0,0.9,0.9} 0.252 &
        \cellcolor[rgb]{0.9,1.0,0.9} 0.171 &
        \cellcolor[rgb]{0.9,0.9,1.0} 0.153 \\
        
        1 & \cellcolor[rgb]{1.0,0.9,0.9} 0.250 &
        \cellcolor[rgb]{0.9,1.0,0.9} 0.226 &
        \cellcolor[rgb]{0.7,0.7,1.0} 0.275 \\
        
        2 & \cellcolor[rgb]{1.0,0.9,0.9} 0.195 &
        \cellcolor[rgb]{0.7,1.0,0.7} 0.914 &
        \cellcolor[rgb]{0.9,0.9,1.0} 0.801 \\
        
        3 & \cellcolor[rgb]{1.0,0.9,0.9} 0.241 &
        \cellcolor[rgb]{0.7,1.0,0.7} 0.944 &
        \cellcolor[rgb]{0.7,0.7,1.0} 1.037 \\

        4 & \cellcolor[rgb]{1.0,0.7,0.7} 0.969 &
        \cellcolor[rgb]{0.9,1.0,0.9} 0.912 &
        \cellcolor[rgb]{0.9,0.9,1.0} 0.194 \\
        
        5 & \cellcolor[rgb]{1.0,0.7,0.7} 0.845 &
        \cellcolor[rgb]{0.9,1.0,0.9} 0.723 &
        \cellcolor[rgb]{0.7,0.7,1.0} 0.936 \\
        
        6 & \cellcolor[rgb]{1.0,0.7,0.7} 0.953 &
        \cellcolor[rgb]{0.7,1.0,0.7} 0.813 &
        \cellcolor[rgb]{0.9,0.9,1.0} 0.173 \\
        
        7 & \cellcolor[rgb]{1.0,0.7,0.7} 0.939 &
        \cellcolor[rgb]{0.7,1.0,0.7} 0.776 &
        \cellcolor[rgb]{0.7,0.7,1.0} 0.885 \\
        
        \hline
    \end{tabular}
    \caption{Input measurements for the Szab\'o model.    A lighter color means
    the simulation was performed with a lower level of 
    the parameter related to that measure, a darker color
    means a higher level of that parameter, see Table~\ref{tab:param-szaboid}. 
    }
    \label{tab:szaboid-cube-result}
\end{table}

Fig.~\ref{fig:snap-szaboid} shows simulation snapshots in limit cases.
Table~\ref{tab:param-szaboid} shows which parameters we vary. 
First, the relaxation time $\tau$, where
low values of $\tau$ favor global alignment. 
The relation between $\tau$ and alignment is indirect, and not explicit; note for instance that the simulation time interval limits the maximum possible alignment.
Second, the 
adhesion parameter $F_{\text{adh}}$: a high value of $F_{\text{adh}}$
favors solid-like behaviour, but it also affects the density and should remain small enough to avoid particle overlap. 
Third, the creation rate: it has to be carefully tuned in 
order to keep a constant density. 
Note that void formation is rare in this model.
Velocity coherence regions are wider than for the  Vicsek model, and the disordered region before the obstacle appears at higher densities (labels 5 and 7).
Here again, at high alignment, particles barely separate and $\Delta$ is high, as in a solid phase.

\subsection{Voronoi model}
 
In a Voronoi model, the degree of freedom is the cell center, but cells have geometrical quantities (a shape, a perimeter, an area, vertices, edges) which can play a role in the dynamics.
The neighbours are defined by the Delaunay triangulation (the dual of the Voronoi tessellation).

We use here the self-propelled particle version of the Voronoi model proposed by Bi {\it et al.}~\cite{Bi2016} and implemented with boundaries and division by Barton {\it et al.}~\cite{Barton2017,SAMoS}.
As in the Szab\'o model, the $i$-th Voronoi velocity is given by an overdamped Langevin equation:
\begin{equation}
\frac{d\vec{r}_i}{dt} = v_0 \hat{n}_i -\mu \nabla_{\vec{r}_i} E 
\end{equation}
where $v_0$ is  the free particle velocity 
and $\hat{n}_i$ the particle
polarity while $\mu$ is the mobility. The last term, $\vec{F}_i = -\nabla_{\vec{r}_i} E$,  is the force term acting  on particle $i$.
It is 
written in terms
of the energy $E$ calculated for the entire Voronoi tiling,
which includes the interaction with neighbors through a preferred area and perimeter:
\begin{equation}
E = \frac{K}{2} \displaystyle\sum_{i} (A_i - A_i^0)^2  +
\frac{\Gamma}{2} \displaystyle\sum_{i} P_i^2  + \displaystyle\sum_{ij} \Lambda  l_{ij}
\end{equation}

Here, the preferred area is $A_i^0$ and each cell's actual
area $A_i$ is determined by its Voronoi tile.
The compressibility modulus $K$ determines the effect of area variation on energy;
$\Gamma$ plays the same role for the perimeter $P_i$, whose preferred value $P^0 = -  \Lambda / \Gamma$ 
is implicit in the last term  of the energy.
The latter is summed over each cell-cell junction ${ij}$, which is a Voronoi edge, and $\Lambda$ is its tension. 

The model can incorporate both an explicit neighbor 
 alignment and the self-persistence of a polarity, with an angular noise, so that the cell polarity evolves according to
 
\begin{equation}
\frac{d\hat{n_i}}{dt} = \vec{\tau}_i + \vec{\xi}_i
\end{equation}

where the 
torque $\tau_i$ acting on the particle is given by
\begin{equation}
\vec{\tau}_i = -\hat{n}_i \times \nabla_{\hat{n}_i} E_{\text{align}}
\end{equation}
 
We separately test  both options (Table~\ref{tab:param-voronoi}).
If $E_{\text{align}}$ is result of the explicit neighbor alignment,
similar to the Vicsek particle model,
then $E_{\text{align}} = -J \sum_j \hat{n}_i \cdot \hat{n}_j$ and in that case $J$
is the alignment parameter.
If $E_{\text{align}}$ is result of the particle self persistence,
similar to the Szab\'o model,
then $E_{\text{align}} = - \frac{1}{\tau} \hat{n}_i \cdot \hat{v}_i$ in which case $\tau$ is the alignment parameter.

The second variable parameter is the cell-cell junction tension $\Lambda$, uniform for all cells and independent of the junction length. This measure maps to the shape parameter $p_0 = - \frac{\Lambda}{\Gamma \sqrt{A_0}}$ that controls the mechanical transition from a rigid to a floppy system in this model~\cite{Bi2016}.
The third parameter is the initial density, $\rho_0$. 
Note that in this model, we create new particles by division:
cells inside the source region divide  every $100$ time steps with a probability
of $3\%$.
When we increase the cell density in the source area, the rate of creation is indirectly increased.
Also, we do not  destroy the cells 
at the end of the channel as this is difficult to integrate into a persistent Delaunay triangulation, so we simply leave enough free space for the particles to migrate. 

\begin{table}[h]
    \centering
    \begin{tabular}{|c|c|c|c|}
            \hline 
            Parameter & Level & Value \\
            \hline
            \hline
            Alignment ($J$) & low  & 0.0  \\
            Alignment ($J$) & high   & 0.5  \\
            Alignment ($\tau$) & low  & 500.0 \\
            Alignment ($\tau$) & high   & 0.5  \\
            Force ($\Lambda$) & low & -4.0 \\
            Force ($\Lambda$) & high  & -4.5 \\
            Creation ($\rho_0$) & low & 1.0 \\
            Creation ($\rho_0$) & high  & 1.5 \\
            \hline
    \end{tabular}
    \caption{
    Limit values for the parameters varied in the Voronoi model. Note the two options for alignment: either $J$ (for neighbours) or $\tau$ (for persistence).
}
    \label{tab:param-voronoi}
\end{table}

The system dimensions in simulation units are: 
channel length 200, width 50, obstacle center position (0, 0), obstacle radius 7.5, 
channel left at $x = -100$, channel right at  $x=100$.
Each cell has an equilibrium area $A_i^0 = \pi$, and stiffness $K =  1$, $\Gamma = 1$ as well as $\mu=1$,  $v_0 = 0.6$ and a rotational noise amplitude $\xi_i(t)\cdot \xi_j(t') = 2 D_r \delta_{ij} \delta(t-t') $ with $2 Dr = 0.5$.


\begin{figure*}[h]
\begin{center}
\includegraphics[width=\textwidth]{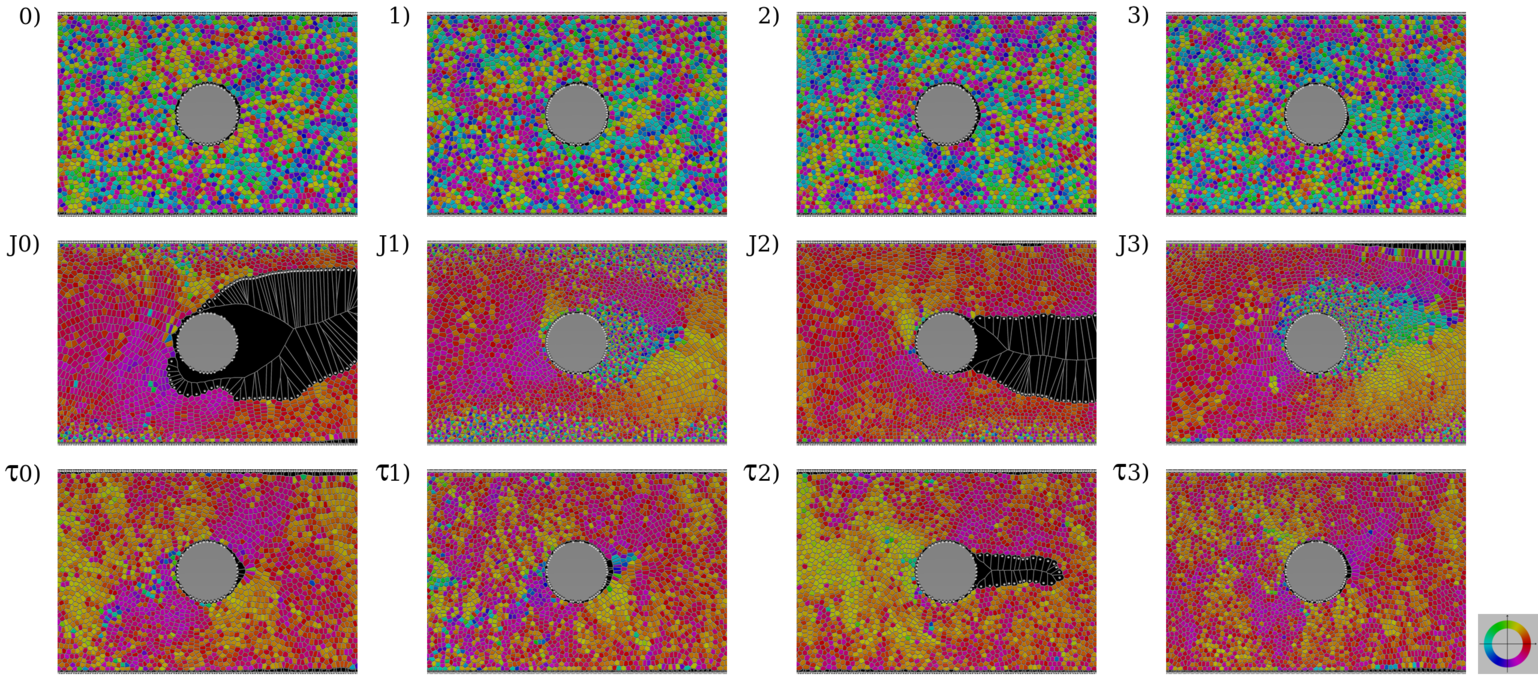}
\caption{
Snapshots for the Voronoi model limit cases.  
The obstacle and the walls are groups of fixed particles represented by small white particles. Clinging particles are also represented this way. Moving particles' color indicate their velocity direction \SH{according to the color scheme at the bottom right}. 
The numbers indicate the levels of each parameter, see 
Tables~\ref{fig:limits-and-cube}a and \ref{tab:param-voronoi}.
The first line corresponds to the case without neighbor alignment and low self-persistence. The second line is for high explicit neighbor alignment, while the third line imposes high self-persistence. 
Images with even numbers present systems with density close to confluence,  while the odd ones are constructed with higher densities. The six images on the left of the figure correspond to low cell-cell junction tension, while the six on the right correspond to high junction tension.
}
\label{fig:snap-voronoi}
\end{center}
\end{figure*}

\begin{table}[h]
    \centering
    \begin{tabular}{|c|c|c|c|}
        \hline 
        id & $\phi$ & $\Delta$ &  $\delta\rho$\\
        \hline
        \hline 
        
        0 & \cellcolor[rgb]{1.0,0.9,0.9} 0.182 &
        \cellcolor[rgb]{0.9,1.0,0.9} 0.373 &
        \cellcolor[rgb]{0.9,0.9,1.0} 0.619 \\
        
        1 & \cellcolor[rgb]{1.0,0.9,0.9} 0.155 &
        \cellcolor[rgb]{0.9,1.0,0.9} 0.011 &
        \cellcolor[rgb]{0.7,0.7,1.0} 0.901 \\
        
        2 & \cellcolor[rgb]{1.0,0.9,0.9} 0.205 &
        \cellcolor[rgb]{0.7,1.0,0.7} 0.624 &
        \cellcolor[rgb]{0.9,0.9,1.0} 0.806 \\
        
        3 & \cellcolor[rgb]{1.0,0.9,0.9} 0.173 &
        \cellcolor[rgb]{0.7,1.0,0.7} 0.061 &
        \cellcolor[rgb]{0.7,0.7,1.0} 1.042 \\

        J0 & \cellcolor[rgb]{0.9,0.75,0.75} 0.582 &
        \cellcolor[rgb]{0.9,1.0,0.9} 0.661 &
        \cellcolor[rgb]{0.9,0.9,1.0} 0.393 \\
        
        J1 & \cellcolor[rgb]{0.9,0.75,0.75} 0.623 &
        \cellcolor[rgb]{0.9,1.0,0.9} 0.666 &
        \cellcolor[rgb]{0.7,0.7,1.0} 1.031 \\
        
        J2 & \cellcolor[rgb]{0.9,0.75,0.75} 0.942 &
        \cellcolor[rgb]{0.7,1.0,0.7} 0.509 &
        \cellcolor[rgb]{0.9,0.9,1.0} 0.507 \\
        
        J3 & \cellcolor[rgb]{0.9,0.75,0.75} 0.903 &
        \cellcolor[rgb]{0.7,1.0,0.7} 0.718 &
        \cellcolor[rgb]{0.7,0.7,1.0} 0.560 \\
    
        $\tau$0 & \cellcolor[rgb]{1.0,0.7,0.7} 0.862 &
        \cellcolor[rgb]{0.9,1.0,0.9} 0.736 &
        \cellcolor[rgb]{0.9,0.9,1.0} 0.621 \\
        
        $\tau$1 & \cellcolor[rgb]{1.0,0.7,0.7} 0.501 &
        \cellcolor[rgb]{0.9,1.0,0.9} 0.099 &
        \cellcolor[rgb]{0.7,0.7,1.0} 1.171 \\
        
        $\tau$2 & \cellcolor[rgb]{1.0,0.7,0.7} 0.903 &
        \cellcolor[rgb]{0.7,1.0,0.7} 0.794 &
        \cellcolor[rgb]{0.9,0.9,1.0} 0.617 \\
        
        $\tau$3 & \cellcolor[rgb]{1.0,0.7,0.7} 0.851 &
        \cellcolor[rgb]{0.7,1.0,0.7} 0.544 &
        \cellcolor[rgb]{0.7,0.7,1.0} 1.114 \\
        
        \hline
    \end{tabular}
        \caption{Input measurements for the Voronoi model.    A lighter color means
    the simulation was performed with a lower level of 
    the parameter related to that measure, a darker color
    means a higher level of that parameter, see Table~\ref{tab:param-voronoi}. 
    }
    \label{tab:voronoi-cube-result}
\end{table}

Fig.~\ref{fig:snap-voronoi} shows simulation snapshots in the limit cases, see Table~\ref{tab:voronoi-cube-result}.
As expected, a low tension favors a liquid behaviour and a high tension favors a solid one. But the density also plays a strong role: high density favors Voronoi topological changes, so that $\Delta$ value is low as in a liquid phase.
The neighbor alignment  is difficult to tune: when we increase the parameter $J$,  before the system reaches a collective behaviour some artifacts
appear.
Examples include the empty spaces after the obstacle when density is low (images J0 and J2), disordered regions after the obstacle when density is high (images J1 and J3), or particle accumulations (top left of image J3).
Conversely, the alignment obtained with high 
self-persistence (bottom row,  indicated by the letter $\tau$) yields more realistic collective behaviours, consistent with the observation of a flocking Voronoi model phase with self-alignment in Refs.~\cite{malinverno2017,giavazzi2018}.

\subsection{Potts model}

In its version without \FG{motil}ity~\cite{Graner1992,Glazier1993,Hirashima2017}, the cellular Potts model represents each cell as a connected set of pixels on a square lattice, like a picture of experimental cells. The degrees of freedom are the cell contours, and each cell has a preferred area. 
The evolution of the pattern is described by the following Monte-Carlo dynamics. At each step, we choose at random a pixel of cell $i$. If it is in the bulk of the cell, it is not examined. If it is near the cell contour, we propose to switch its value by copying that of a neighboring pixel, in another  cell $j$. The energy cost $\Delta H$ (where $H$ is the total system energy) that this copy incurs is evaluated. If the energy $H$ of the system would decrease with the proposed copy, then it is 
always accepted (which is equivalent to moving the junction between $i$ and $j$ by one pixel). If the energy would instead increase, the proposed copy will be
accepted with a probability that exponentially depends on the cost $\Delta H$  and on a fluctuation allowance, $\beta$:
\begin{eqnarray}
P = \left\{ \begin{array}{cl} e^{- \beta \Delta H} &;\; \Delta H
> 0 \\ \ 1 &;\; \Delta H \le 0
\end{array} \right.
\end{eqnarray}
In this case without any cell \FG{motil}ity, the energy is given by
\begin{equation}
H = \displaystyle\sum_{i\sim j} J 
+ \lambda_A \displaystyle\sum_{i} \left(A_i - A_{0}\right)^2, 
\end{equation}
where the first term is the tension of the junction between cells $i$ and $j$, and the sum is performed over all pixels at the junction, hence encompasses the junction length.
 As $J$ increases, 
changes are less probable, and the tissue has a more solid-like behaviour.
The second term is the area conservation: the cell has an
equilibrium area $A_0$ and an actual area $A_i$, while the parameter $\lambda_A$
is an area compressibility modulus. When there is a free space between cells, it is treated as a zone with no preferred area and no compression modulus, and its border with a cell has tension $J$.
The channel walls and the obstacle are treated as a fixed zone without changes.

In the present work, we add \FG{motil}ity to the Potts model \CB{based on Käfer {\it et al.}~\cite{Kafer2006}} cells by introducing the following \FG{motil}e force $ \vec{F}$:
\begin{equation}
\Delta H =  \vec{F_i} \cdot \vec{c}(i,j),
\end{equation}
where $\vec{c}$ is the copy vector. That is, for each pixel copy 
proposed during the Monte-Carlo step, the vector $\vec{c}$ which links both pixels is a proxy of the direction of movement for the whole cell. 
If the copy is aligned with the force, the energy decreases and
the copy is favored; conversely, if the copy vector has a  direction opposed
to the force, the energy increases 
and the copy is less probable. 
If the copy is perpendicular, it does not change the energy:
hence some random perpendicular copies occur.

The \FG{motil}e force is:
\begin{equation}
    \vec{F_i}(t+\Delta t)  = \alpha \hat{P_i}(t)
\end{equation} 
Here $\alpha$ is the total \FG{motil}ity parameter;
if it is zero the cell has no \FG{motil}ity and $\overrightarrow {v_i}$ is the past cell velocity before the change.
The cell polarity is 
defined as the direction of the \FG{motil}e force, and thus of $\vec{c}(i,j)$. Both neighbor alignment and self-persistence terms can be implemented as:

\begin{equation}
    \hat{P_i}(t+\Delta t)  = \left[ 
    \lambda_C   \displaystyle\sum_{j\sim i} \hat{P}_j(t) + \lambda_P \hat{v}_i(t)
    \right].
    \label{eq:potts_polarity}
\end{equation}

We have observed that, since $\vec{c}(i,j)$ is pixelated, it yields highly fluctuating simulations. 
We thus mostly study the neighbor alignment term, by making $  \lambda_C$ variable.
 The second variable parameter is $J$, i.e. the
tension of cell-cell junctions. The third parameter is  the division   area $A^*$. 
In the source region, the mother cells grow, and once they reach $A^*$ they divide into two particles.
To obtain higher density values
 we decrease $A^*$ (Table~\ref{tab:param-potts}), and other parameters are fixed. 

\begin{table}[h]
    \centering
    \begin{tabular}{|c|c|c|c|}
            \hline 
            Parameter & Level & Value \\
            \hline
            \hline
            Alignment $(\lambda_C)$ & low  & 0.0  \\
            Alignment $(\lambda_C)$ & high   & 5.0  \\
            Force $(J)$ & low  & 50 \\
            Force $(J)$ & high   & 150 \\
            Creation $(A^*)$ & low & 80 \\
            Creation $(A^*)$ & high & 53.33 \\
            \hline
    \end{tabular}
    \caption{Limit values for the parameters varied in the Potts model. Note that bigger area $A^*$ means less divisions. 
    }
    \label{tab:param-potts}
\end{table}

All dimensions are expressed in pixels:
channel length 2020, width 520, obstacle center position (810,260), obstacle radius 74, 
cell target area $A_0 = 100$.
We use $\alpha = 100$ and $\beta=1/50$, $\lambda_A = 10$, $\lambda_P  = 1$.

\begin{figure*}[h]
\begin{center}
\includegraphics[width=\textwidth]{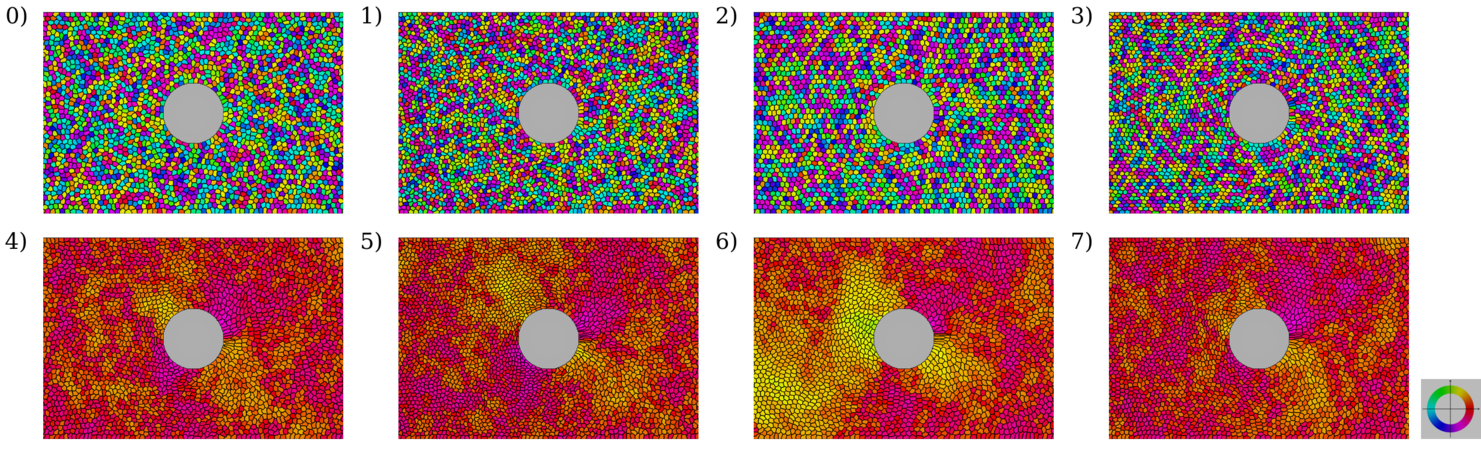}
\caption{
Snapshots for the Potts model's eight limit cases; same caption as 
Fig.~\ref{fig:1-particle-simul}. See Table~\ref{tab:param-potts} for parameter values.}
\label{fig:snap-potts}
\end{center}
\end{figure*}

\begin{table}[h]
    \centering
    \begin{tabular}{|c|c|c|c|}
        \hline 
        id & $\phi$ & $\Delta$ &  $\delta\rho$\\
        \hline
        \hline 
        0 & \cellcolor[rgb]{1.0,0.9,0.9} 0.101 &
        \cellcolor[rgb]{0.9,1.0,0.9} 0.090 &
        \cellcolor[rgb]{0.9,0.9,1.0} 0.224 \\
        
        1 & \cellcolor[rgb]{1.0,0.9,0.9} 0.084 &
        \cellcolor[rgb]{0.9,1.0,0.9} 0.099 &
        \cellcolor[rgb]{0.7,0.7,1.0} 0.654 \\
        
        2 & \cellcolor[rgb]{1.0,0.9,0.9} 0.100 &
        \cellcolor[rgb]{0.7,1.0,0.7} 0.777 &
        \cellcolor[rgb]{0.9,0.9,1.0} 0.282 \\
        
        3 & \cellcolor[rgb]{1.0,0.9,0.9} 0.089 &
        \cellcolor[rgb]{0.7,1.0,0.7} 0.822 &
        \cellcolor[rgb]{0.7,0.7,1.0} 0.797 \\
 
        4 & \cellcolor[rgb]{1.0,0.7,0.7} 0.498 &
        \cellcolor[rgb]{0.9,1.0,0.9} 0.772 &
        \cellcolor[rgb]{0.9,0.9,1.0} 0.187 \\
        
        5 & \cellcolor[rgb]{1.0,0.7,0.7} 0.460 &
        \cellcolor[rgb]{0.9,1.0,0.9} 0.613 &
        \cellcolor[rgb]{0.7,0.7,1.0} 1.114 \\
        
        6 & \cellcolor[rgb]{1.0,0.7,0.7} 0.485 &
        \cellcolor[rgb]{0.7,1.0,0.7} 0.791 &
        \cellcolor[rgb]{0.9,0.9,1.0} 0.058 \\
        
        7 & \cellcolor[rgb]{1.0,0.7,0.7} 0.570 &
        \cellcolor[rgb]{0.7,1.0,0.7} 0.811 &
        \cellcolor[rgb]{0.7,0.7,1.0} 1.114 \\
        
        \hline
    \end{tabular}
        \caption{Input measurements for the Potts model.    A lighter color means
    the simulation was performed with a lower level of 
    the parameter related to that measure, a darker color
    means a higher level of that parameter, see Table~\ref{tab:param-potts}. 
    }
   \label{tab:potts-cube-result}
\end{table}
  
 Fig.~\ref{fig:snap-potts} shows simulation snapshots in limit cases, see Table~\ref{tab:potts-cube-result}.
While voids between cells are possible to simulate, here we do not intend to simulate them so by construction there are none.
Note that the polarization is nearly random in the top images where there is no collective motion. When in collective motion, the polarization is overall aligned, with direction fluctuations only close to the obstacle.

\subsection{Multiparticle model}

In this work we introduce a Multiparticle model  where several vertices are free to move and interacting pairwise, in the same spirit as Refs.~\cite{Teixeira2021,Treado2021}.

\begin{figure}[h!]
    \centering
    \includegraphics[width=8cm]{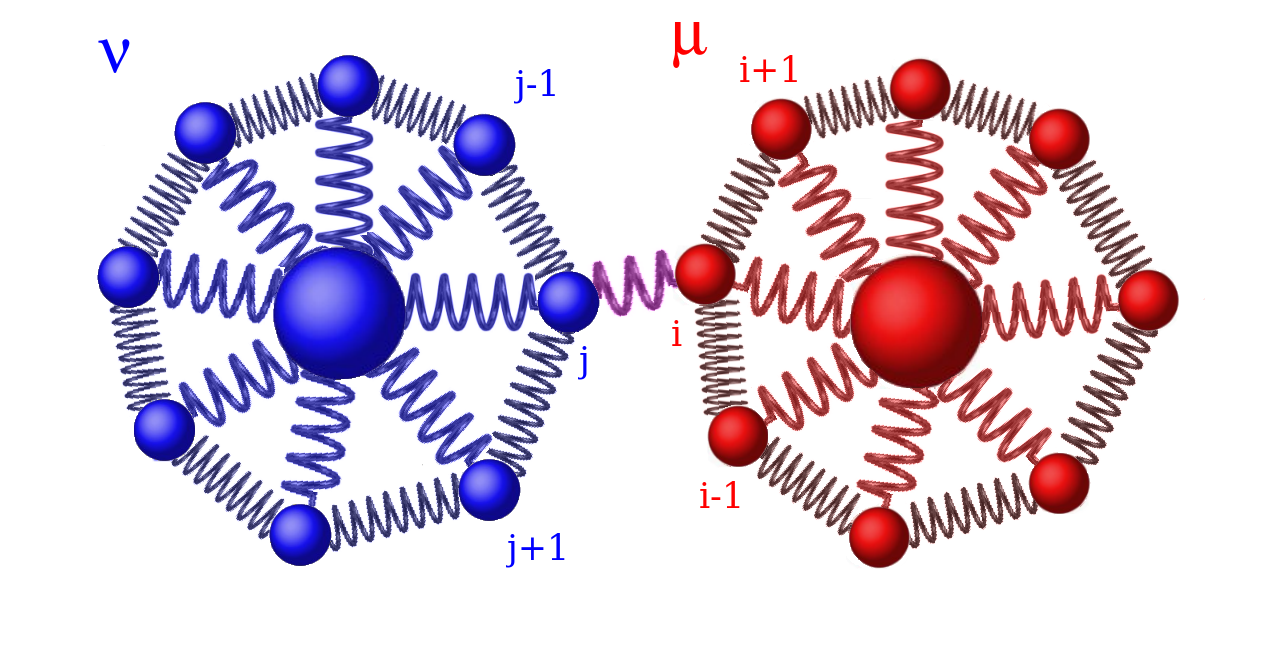}
    \caption{Multiparticle model. Schema of the springs
    composing the extended cell model, including interaction
    with neighboring cells. Each cell  and its
    central particle is labelled by a Greek letter, here $\mu$ 
    and $\nu$. The cells $\mu$ 
    and $\nu$ interact only via their peripheral particles
   ${\mu,i}$ and ${\nu,j}$. }
    \label{fig:my_label}
\end{figure}

Each cell is composed of two kinds of \FG{motil}e 
particles: a central one and several peripheral ones (Fig.~\ref{fig:my_label}). 
The central one (representing the nucleus), also labelled $\mu$,  interacts only
with the peripheral particles of the same cell 
(representing the membrane or the cytoskeleton), which are labelled ${\mu,i}$.  

Within a given cell
the neighborhood is fixed. 
Each peripheral 
particle is always a 
neighbor to the  central particle, and to two other peripheral particles.
In addition, the peripheral particles of one cell are capable
of interacting with the peripheral ones from neighboring
cells and so are responsible
for cell-cell interactions. 
We use the following notations: $v^\mu$ is the speed of 
the central particle   of cell $\mu$; $v_i^\mu$ is the speed of the peripheral particle $i$ of cell  ${\mu}$;
when two particles are neighbors we note $i \sim j$, finally  $\vec{r}_{i,j}^{\;\mu,\nu}$
is the vector connecting the position of particle $i$ from cell
$\mu$ to the position of particle $j$ from cell $\nu$.

An individual particle is described by an adapted Vicsek
equation and the central particle movement of the $\mu$-th cell is given by:
\begin{equation}
\vec{x}^\mu (t+\Delta t) = \vec{x}^\mu (t) + \vec{v}^\mu (t) \Delta t .
\end{equation}
As in the Vicsek model, the velocity has a constant modulus $|\vec{v}^{\mu}|$. Its
 direction depends on the alignment and forces
of the peripheral particles of the same cell.
The velocity direction $\theta^{\mu}$ of the central particle evolves according to

\begin{equation}
    \theta^\mu (t+\Delta t) = \arg \left[\alpha \sum_{i \in \mu} \vec{v}_i(t) + \beta \sum_{i \in \mu} \vec{h}_{i}(\vec{r}_i,t)+ \eta \vec{u}(t)\right],
\end{equation}
where $\alpha$, $\beta$ and $\eta$, respectively,  regulate the weights of the alignment with the peripheral particles velocity, the harmonic forces, $\vec{h}_i$, produced by peripheral particles on the central one, and the intensity of the unitary noise vector $\vec{u}$.  
The evolution equation for the peripheral particle 
$i$ in cell ${\mu}$ is similar
\begin{equation}
\vec{x}^{\mu}_i(t+\Delta t) = \vec{x}^{\mu}_i(t) + \vec{v}^{\mu}_i(t)\Delta t. \end{equation}

The interaction between peripheral particles  of the same cell and particles of  different cells ($i$ and $j$ in Fig.~\ref{fig:my_label}) results from the sum of several contributions. The particle ${i}$ in cell $\mu$ has velocity direction $\theta$ given by
\begin{equation}
\theta_{i}^{\mu} =  \arg  \left[ A_i^\mu  +   F_i^\mu   + G_i^\mu    + H_i^\mu    +   T_i^\mu \right],
\label{theta_multi}
\end{equation}
where each term is explained one by one below.

First, consider a peripheral particle $i$
that is part of the cell $\mu$.  
The total alignment acting on it, $A^{\mu}_i$,
is composed of the central particle direction 
$\hat{v}_\mu$, that is a self-persistence term, and the direction of the velocity of neighboring  
peripheral particles either from cell $\mu$ and from neighboring cells $\nu$:

\begin{equation}
A^{\mu}_i =  \alpha \hat{v}^\mu + \alpha \sum_{i,j \in \mu} \hat{v}^{\mu,\mu}_{i,j} + \alpha_1 \sum_{i \in \mu ,j \in \nu}  \hat{v}^{\mu,\nu}_{i,j}.
\end{equation}

The second term in Eq.~(\ref{theta_multi}) is a force term and also involves contributions from the central particle and from peripheral particles,

\begin{equation}
F^{\mu}_i =  \beta \left(\vec{h}^{\mu}_i + \sum_{ j\sim i}  \vec{h}^{\mu}_{i,j} \right) + \beta_1 \sum_{ i \in \mu ,j \in \nu} \vec{f}^{\mu,\nu}_{i,j}
\end{equation}
where $\vec{h}$
is an infinite range harmonic interaction between peripheral particles of the same cell or with the central particle of their cell. The last term represents interactions with peripheral particles of the same cell $\mu$ when not first neighbors, or from a neighbor cell $\nu$. This last force between pairs is inspired by the force for
Vicsek-like particles (Eq.~\ref{eq:vicsek_force}): it is radial, with limited reach
and its module depends 
on the distance $r_{i,j}^{\mu,\nu}$ between 
peripheral particles

\begin{eqnarray}
f(r_{i,j}^{\mu,\nu}) = \left\{ \begin{array}{cl} 0 & \;\;\;\;\;\;\;\;  r_{i,j}^{\mu,\nu} \ge r_{\text{max}} \\ 1 -
\frac{r_{i,j}^{\mu,\nu}}{r_{\text{eq}}} & \;\;\;\;\;\;\;\;  r_{c} < r_{i,j}^{\mu,\nu} < r_{\text{max}} \\ f_c & \;\;\;\;\;\;\;\; 
r_{i,j}^{\mu,\nu} \le r_{c}. \end{array} \right. 
\label{eq:superboid_force_radial}
\end{eqnarray}
Here, $r_{\text{max}}$ is the cut-off, or maximum interaction distance, $r_{\text{eq}}$ is the 
equilibrium distance, $r_{c}$ is the core size, and $f_c$ plays the role of an infinite repulsion force. In practice, in the simulation it is set to a large value compared to typical forces in the system.

The next force reflects the  cell area constraint:
\begin{equation}
G^{\mu}_i = -k_A(A^\mu-A_0) \hat{r}^\mu_i, 
\label{eq:superboid_force_area}
\end{equation}
where $A^\mu$ is the instantaneous cell area, $A_0$ is a target area, $\hat{r}^\mu_i$ is a unitary radial vector and $k_A$ is a stiffness constant.

The polygonal shape of each cell is not impenetrable:
in principle a peripheral particle could invade another cell. In practice this seldom happens, but for these rare cases we introduce a force to repel the invader: 
\begin{eqnarray}
H^{\mu}_j = \left\{ \begin{array}{cl} f_{c}
\hat{r}_i^{\mu,\nu}&\;\; {\mathrm{if} } \;  j\; {\mathrm{inside} }\; \mu
 \\  0 &\;\; {\mathrm{else,}} 
\end{array} \right.
\end{eqnarray}
with $f_c$ and $\hat{r}_i$ as defined above, and with an equal force with opposite sign that is applied to the center particle of cell $\mu$. 

Since the topological relations between peripheral particles are fixed within a cell, we introduce a torque that keeps the particle near the
correct relative angle with its neighbors. The tangential force resulting from this torque is given by

\begin{equation}
T^{\mu}_i = \kappa r^\mu_i \sum_{j=\pm 1} \phi^{\mu}_j  - \phi_0, 
\end{equation}
where $\phi_0$ is an equilibrium angle, $r_i^\mu $ is the radial distance to the center particle,  $\phi_{i\pm 1}^\mu$ is the angle between peripheral particles $i$ and $i\pm 1$, and $\kappa$ is a constant. 

In this work we keep constant all parameters (Table \ref{tab:param-super-constant}) except for three parameters we vary (Table~\ref{tab:param-super}).

\begin{table}[h]
    \centering
    \begin{tabular}{|c|c||}
            \hline 
            Parameter & Value \\
            \hline
            \hline
            $N$ & 20 \\
            $\alpha$ & 14\\
            $\beta$ & 1\\
            $\eta$ & 1\\
            $r_{eq}$ & 1.1 \\
            $r_{max}$ & 1.3 \\
            $k_a$& 10 \\
            $\kappa$ & 10 \\
            $\phi_0$ & 2$\pi/N$\\
            $R$  & $N/(2\pi)$\\
            $A_0$ & $\pi R^2$     \\
            \hline
    \end{tabular}
    \caption{Parameters kept constant in the Multiparticle  model. $N$ is the number of peripheral particles composing each cell.}
    \label{tab:param-super-constant}
\end{table}

\begin{table}[h]
    \centering
    \begin{tabular}{|c|c|c|c|}
            \hline 
            Parameter & Level & Value \\
            \hline
            \hline
            Alignment $(\alpha_{1})$ & low  & 0.0 \\
            Alignment $(\alpha_{1})$ & high   & 14.0  \\
            Force $(\beta_1)$ & low  & 1.0 \\
            Force $(\beta_1)$ & high & 2.5 \\
            Creation $(\tau)$ & low & 50 \\
            Creation $(\tau)$ & high  & 30 \\
            \hline
    \end{tabular}
    \caption{Limit values for the parameters varied in the Multiparticle  model.}
    \label{tab:param-super}
\end{table}

\begin{figure*}[h]
\begin{center}
\includegraphics[width=\textwidth]{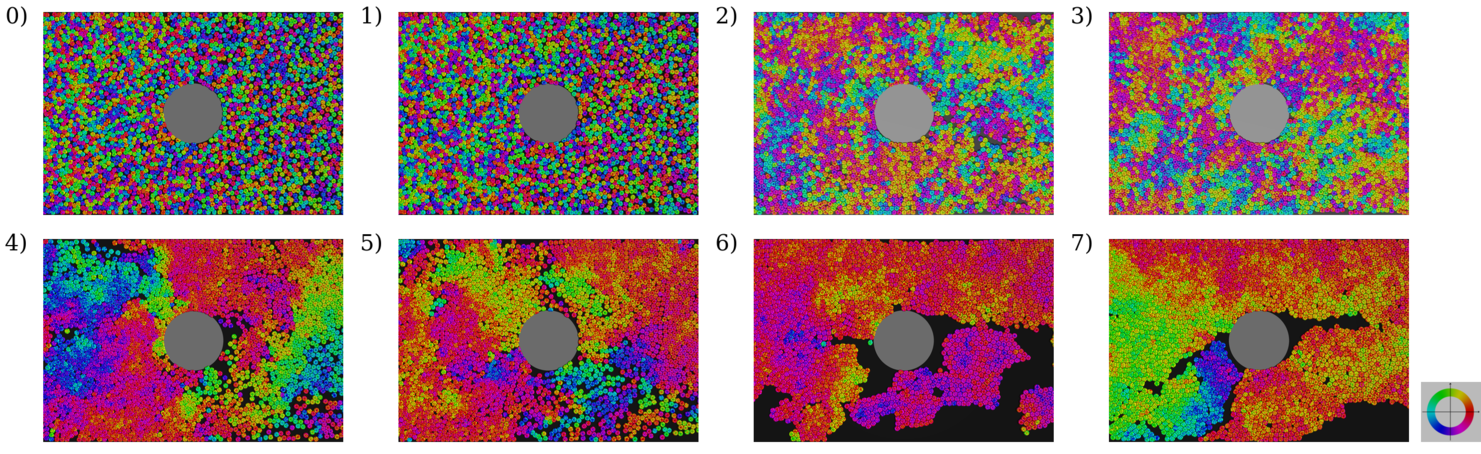}
\caption{
Snapshots for the Multiparticle's eight limit cases; same caption as 
Fig.~\ref{fig:1-particle-simul}. See Table~\ref{tab:param-super} for parameter values.
}
\label{fig:5-multi-simul}
\end{center}
\end{figure*}

\begin{table}[h]
    \centering
    \begin{tabular}{|c|c|c|c|}
        \hline 
        id & $\phi$ & $\Delta$ &  $\delta\rho$\\
        \hline
        \hline 
        0 & \cellcolor[rgb]{1.0,0.9,0.9} 0.182 &
        \cellcolor[rgb]{0.9,1.0,0.9} 0.086 &
        \cellcolor[rgb]{0.9,0.9,1.0} 0.098 \\
        
        1 & \cellcolor[rgb]{1.0,0.9,0.9} 0.161 &
        \cellcolor[rgb]{0.9,1.0,0.9} 0.145 &
        \cellcolor[rgb]{0.7,0.7,1.0} -0.013 \\
        
        2 & \cellcolor[rgb]{1.0,0.9,0.9} 0.285 &
        \cellcolor[rgb]{0.7,1.0,0.7} 0.337 &
        \cellcolor[rgb]{0.9,0.9,1.0} -0.050 \\
        
        3 & \cellcolor[rgb]{1.0,0.9,0.9} 0.290 &
        \cellcolor[rgb]{0.7,1.0,0.7} 0.353 &
        \cellcolor[rgb]{0.7,0.7,1.0} -0.010 \\
 
        4 & \cellcolor[rgb]{1.0,0.7,0.7} 0.643 &
        \cellcolor[rgb]{0.9,1.0,0.9} 0.439 &
        \cellcolor[rgb]{0.9,0.9,1.0} -0.028 \\
        
        5 & \cellcolor[rgb]{1.0,0.7,0.7} 0.599 &
        \cellcolor[rgb]{0.9,1.0,0.9} 0.654 &
        \cellcolor[rgb]{0.7,0.7,1.0} 0.030 \\
        
        6 & \cellcolor[rgb]{1.0,0.7,0.7} 0.671 &
        \cellcolor[rgb]{0.7,1.0,0.7} 0.720 &
        \cellcolor[rgb]{0.9,0.9,1.0} -0.240 \\
        
        7 & \cellcolor[rgb]{1.0,0.7,0.7} 0.794 &
        \cellcolor[rgb]{0.7,1.0,0.7} 0.708 &
        \cellcolor[rgb]{0.7,0.7,1.0} -0.079 \\
        
        \hline
    \end{tabular}
    \caption{Input measurements for the Multiparticle model.    A lighter color means
    the simulation was performed with a lower level of 
    the parameter related to that measure, a darker color
    means a higher level of that parameter, see Table~\ref{tab:param-super}. 
    }
    \label{tab:multiparticle-cube-result}
\end{table}

The first parameter we vary is the external alignment 
$\alpha_1$, 
which we increase in order to establish collective movement.

The second parameter we vary is the attractive force between different cells $\beta_1$. If 
the attractive force $\beta_1$ is low or even zero, particles
from different cells still repel each other due to core repulsion. 
All  forces are fixed at a value carefully chosen in
order to prevent artifacts such as cell breakage, overlap or collapse. 

The third parameter is the cell creation rate, which determines the density. 
As in the Potts model, the creation of new particles is implemented by cell division, which happens at a given rate, $\tau$.


Fig.~\ref{fig:5-multi-simul} shows simulation snapshots in limit cases, see Table~\ref{tab:multiparticle-cube-result}.
Note the presence of voids and coherent polarization patches when the alignment is high and the motion is collective. 
Even with a high self-persistence value, collective alignment is never reached, probably because peripheral particles generate a lot of noise.

\begin{figure*}[h]
\begin{center}
\includegraphics[width=\textwidth]{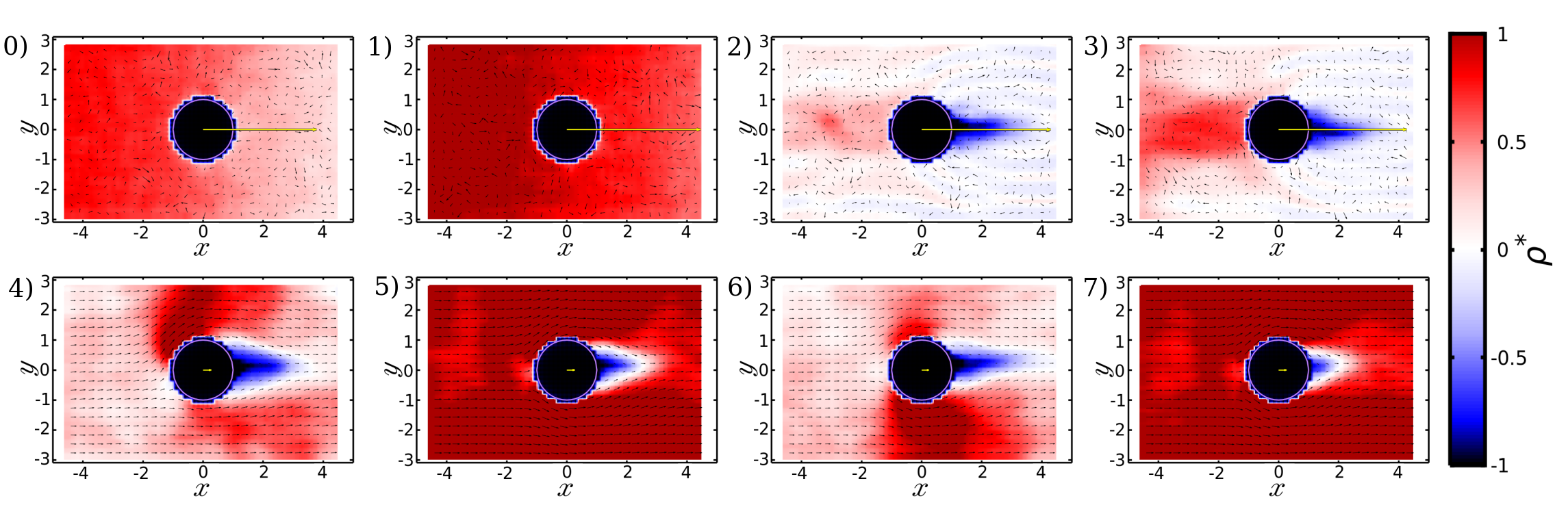}
\caption{Vicsek model: density and velocity. The numbers in the image are the labels detailed in Table~\ref{tab:cubo}. 
The values for the parameters used in this model are specified in Table~\ref{tab:param-boid}.
Images with even numbers present systems with density close to confluence, 
while the odd ones are constructed with higher densities. 
The top row presents the low alignment cases while the bottom
one presents the high alignment ones. The four images on the left correspond to low  forces (liquid-like), while the 
four images on the right correspond to high  forces 
(solid-like). 
The images are restricted to an area around the obstacle; particle source and sink regions are not depicted. }
\label{fig:den_eq_vel_vicsek}
\end{center}
\end{figure*}

\begin{figure*}[h]
\begin{center}
\includegraphics[width=\textwidth]{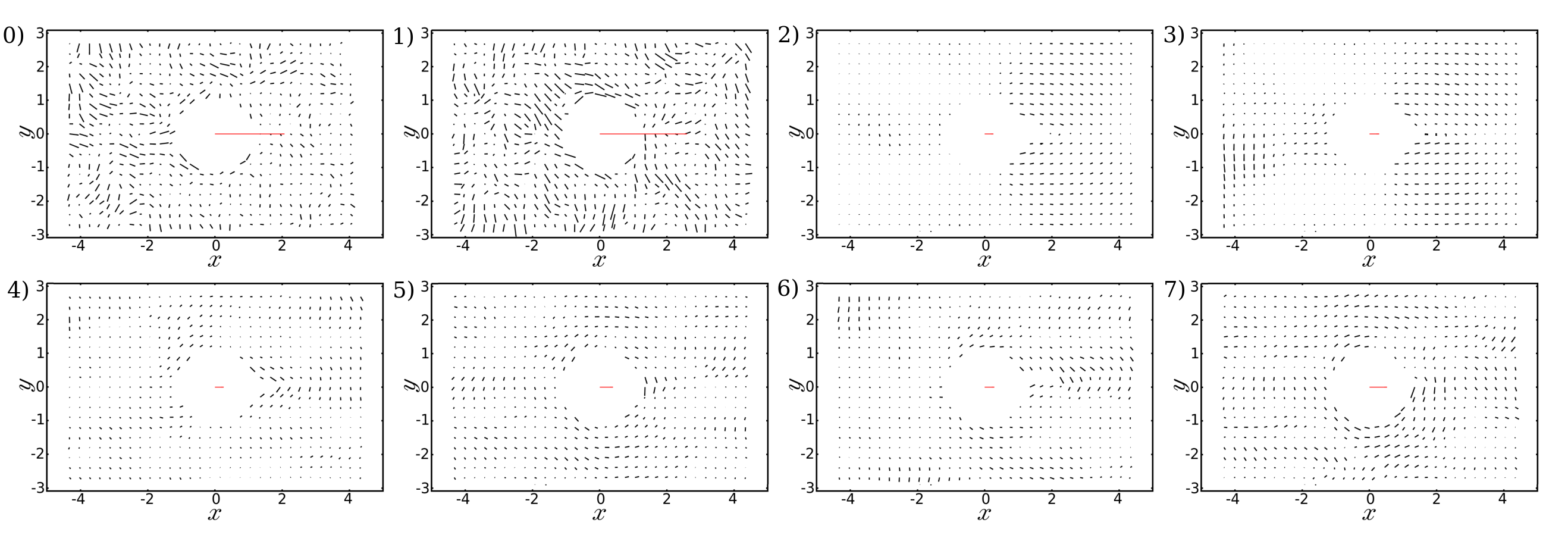}
\caption{ Vicsek model: deformation anisotropy, for the same data as in Fig. ~\ref{fig:den_eq_vel_vicsek}.}
\label{fig:deform_vel_vicsek}
\end{center}
\end{figure*}

\begin{figure*}[h]
\begin{center}
\includegraphics[width=\textwidth]{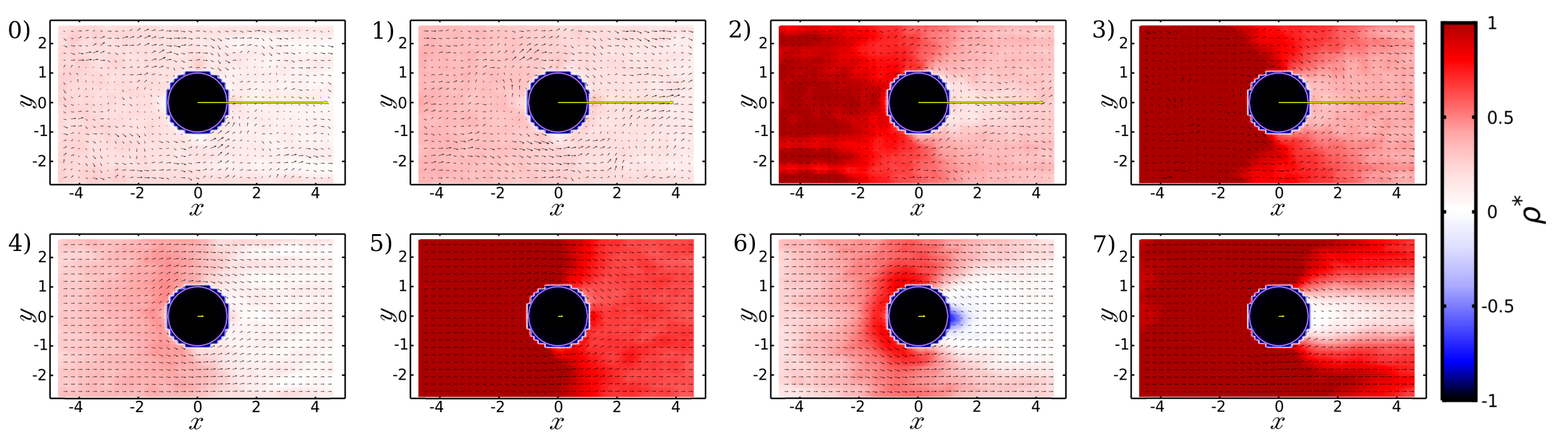}
\caption{Szab\'o model: density and velocity, for the 8 cases outlined in Tables~\ref{tab:param-szaboid} and \ref{tab:szaboid-cube-result}. }
\label{fig:den_eq_vel_szabo}
\end{center}
\end{figure*}

\begin{figure*}[h]
\begin{center}
\includegraphics[width=\textwidth]{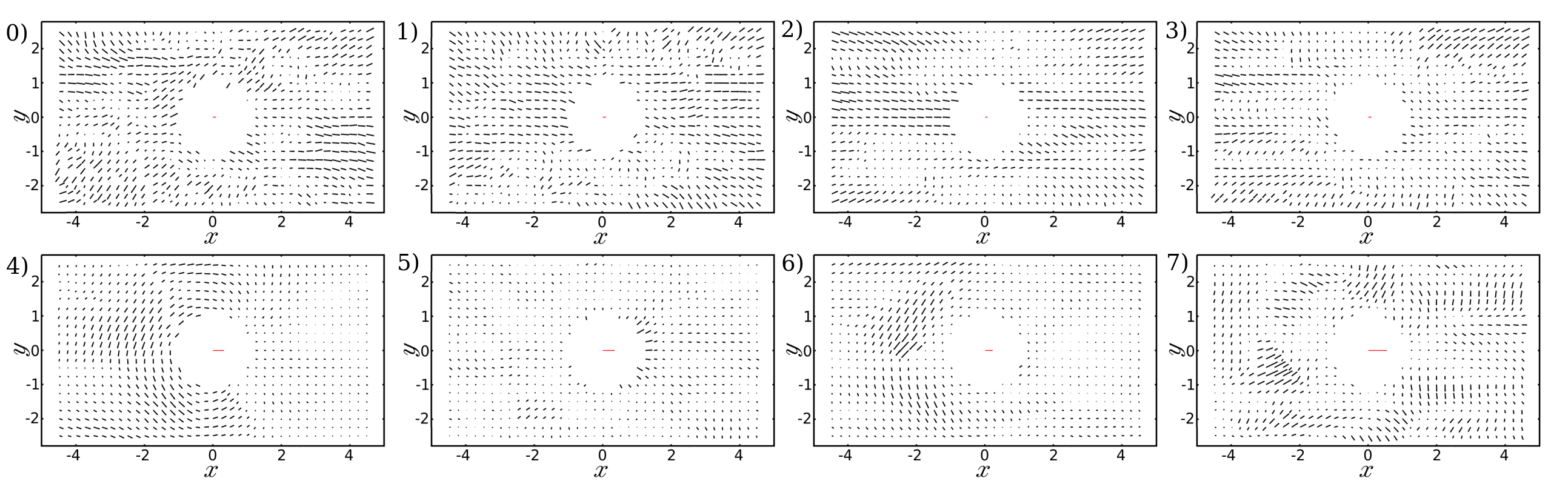}
\caption{Szab\'o model: deformation anisotropy, for the 8 cases outlined in Tables~\ref{tab:param-szaboid} and \ref{tab:szaboid-cube-result}.}
\label{fig:deform_vel_szabo}
\end{center}
\end{figure*}

\begin{figure*}[h]
\begin{center}
\includegraphics[width=\textwidth]{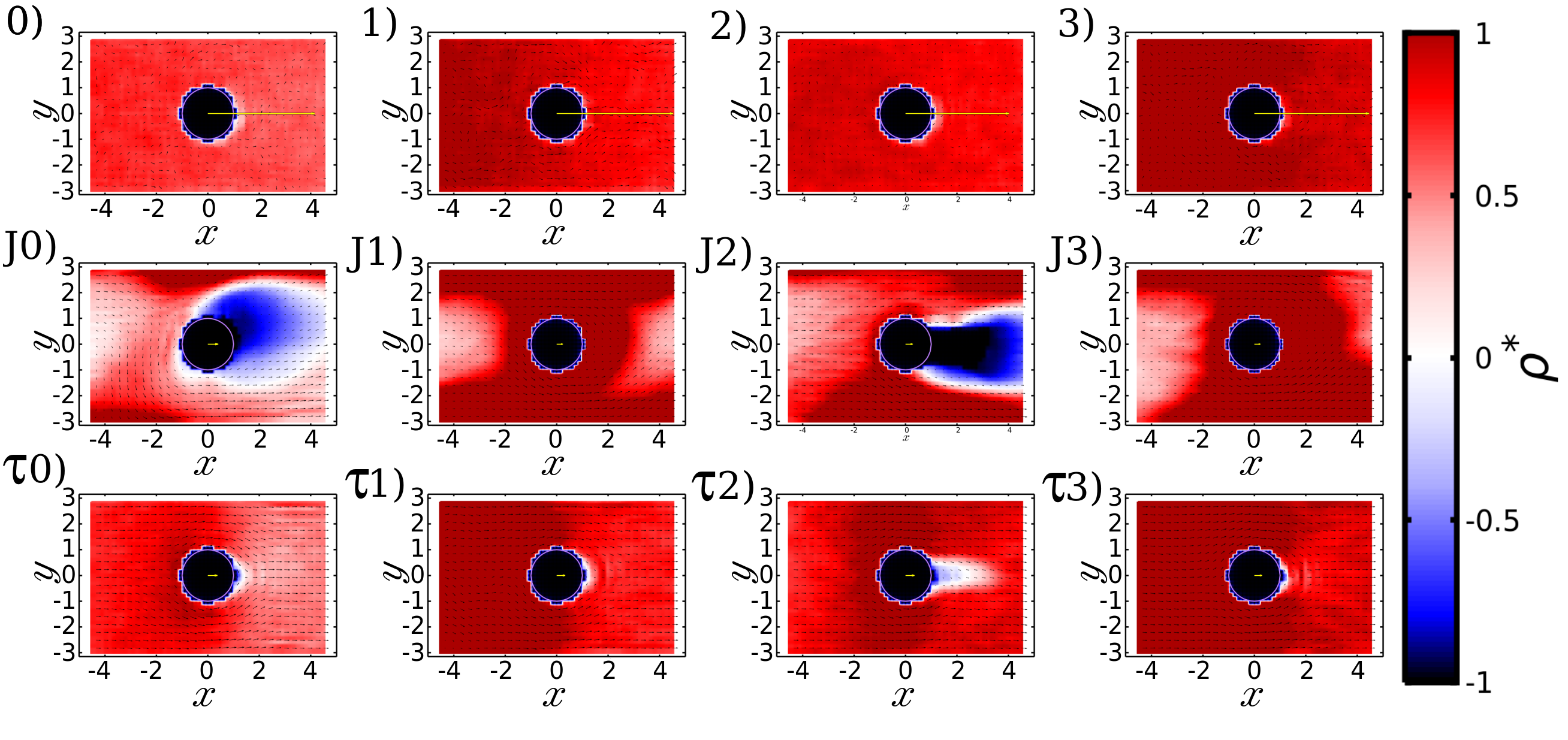}
\caption{Voronoi model: density and velocity, for the 12 cases outlined in Tables~\ref{tab:param-voronoi} and \ref{tab:voronoi-cube-result}. In J2, regions downstream of the obstacle with aberrant cell shapes and velocities have been removed.}
\label{fig:den_eq_vel_voronoi}
\end{center}
\end{figure*}

\begin{figure*}[h]
\begin{center}
\includegraphics[width=\textwidth]{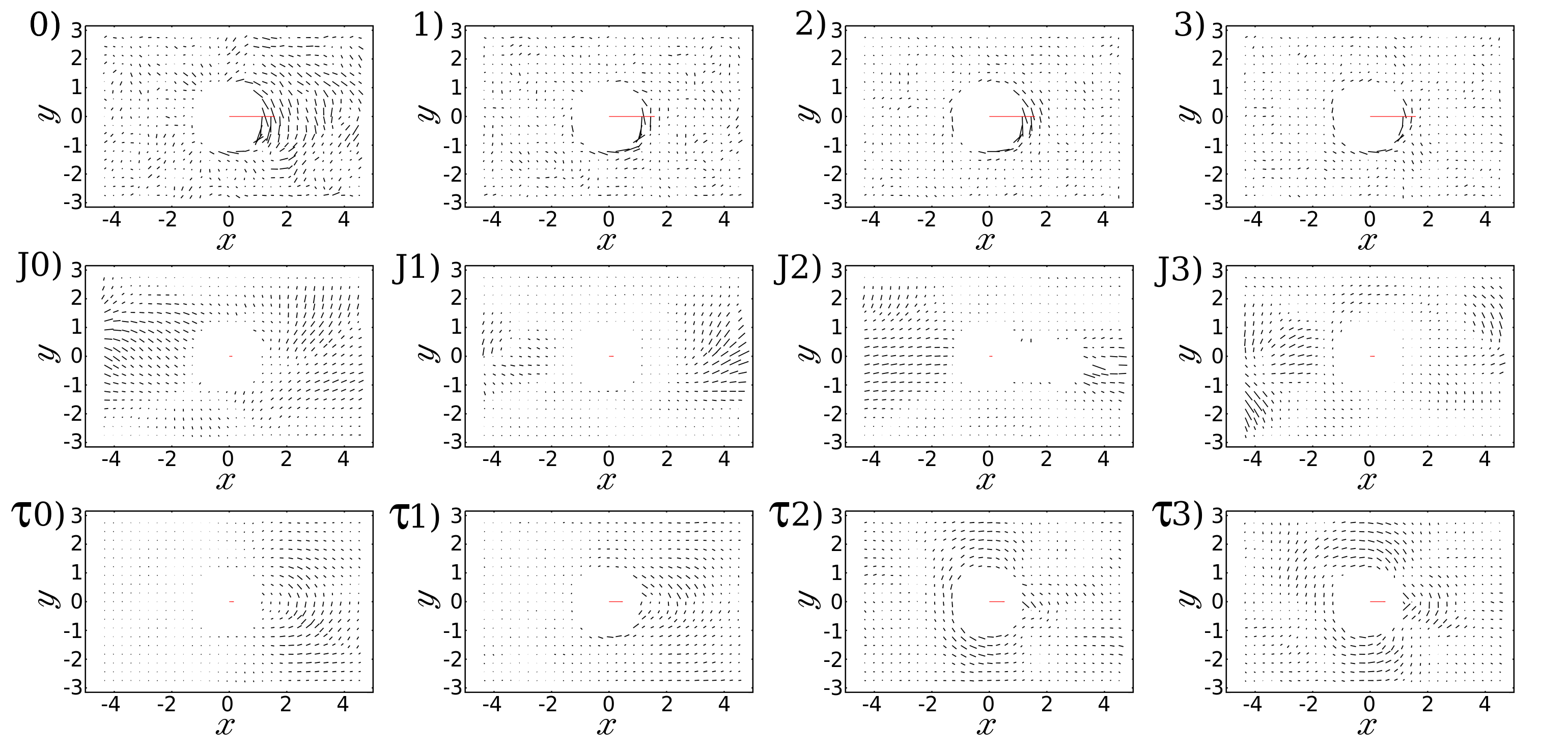}
\caption{Pictures/Voronoi model: deformation anisotropy, for the 12 cases outlined in Tables~\ref{tab:param-voronoi} and \ref{tab:voronoi-cube-result}.  In J2, regions downstream of the obstacle  with aberrant cell shapes and velocities have been removed.}
\label{fig:deform_voronoi}
\end{center}
\end{figure*}

\begin{figure*}[h]
\begin{center}
\includegraphics[width=\textwidth]{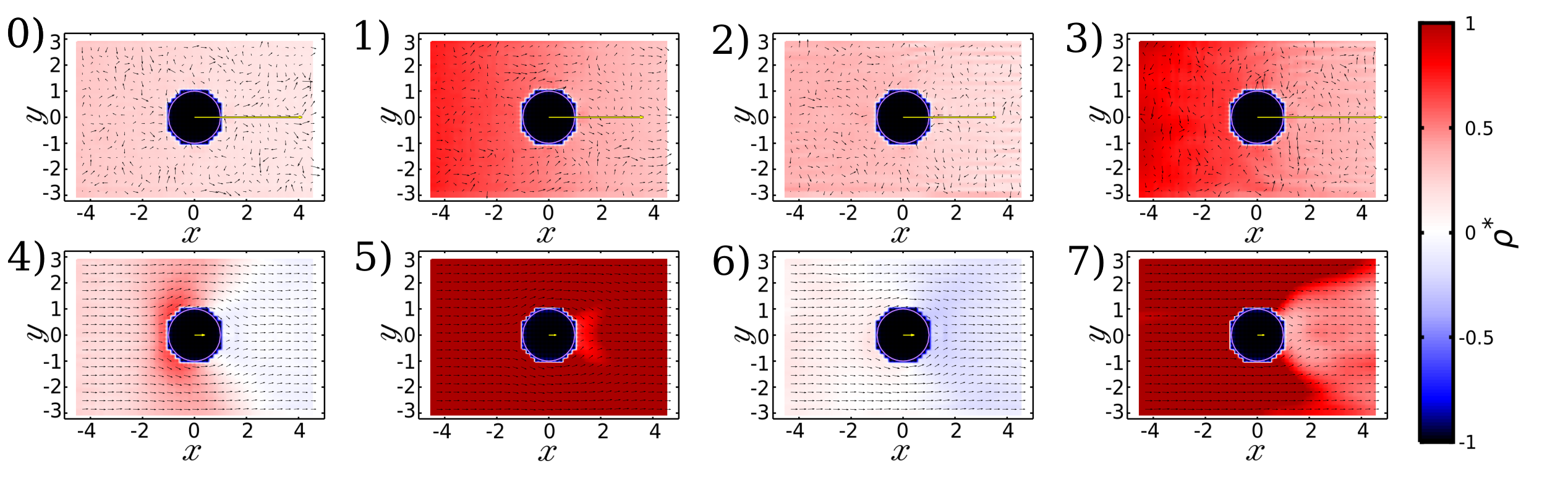}
\caption{Potts model: density and velocity, for the 8 cases outlined in Tables~\ref{tab:param-potts} and \ref{tab:potts-cube-result}.}
\label{fig:den_eq_vel_potts}
\end{center}
\end{figure*}

\begin{figure*}[h]
\begin{center}
\includegraphics[width=\textwidth]{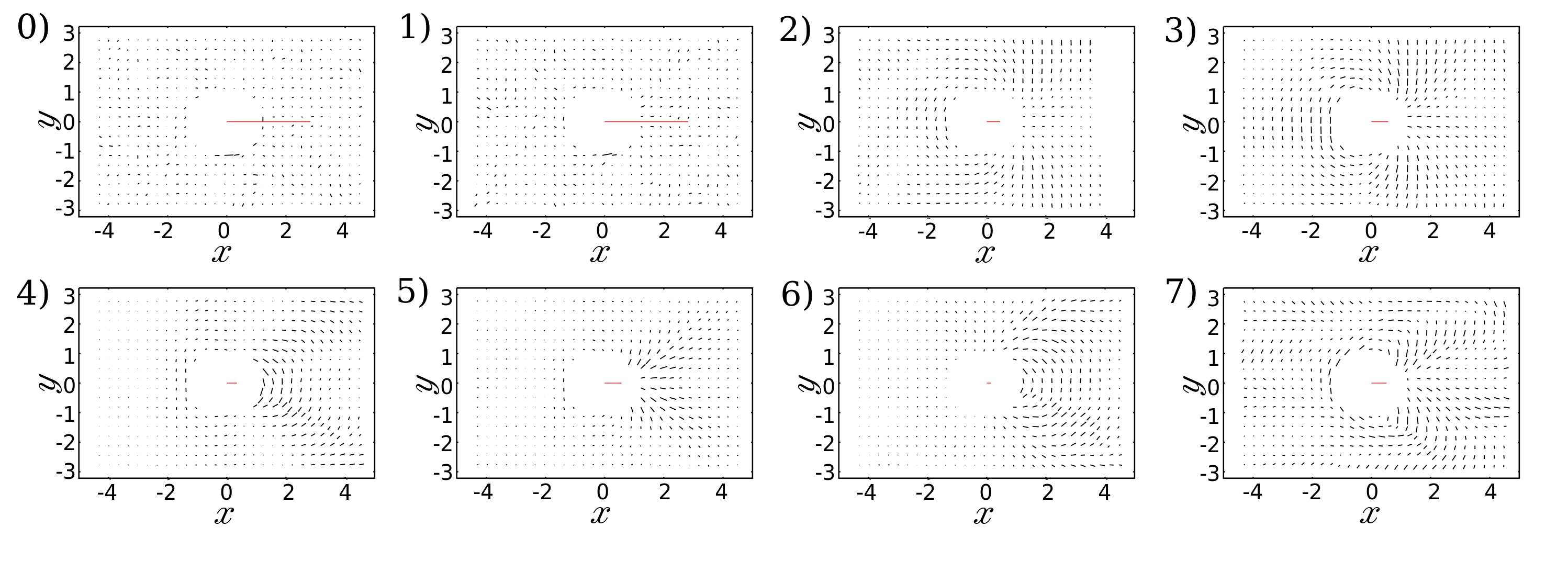}
\caption{Potts model: deformation anisotropy, for the 8 cases outlined in Tables~\ref{tab:param-potts} and \ref{tab:potts-cube-result}.}
\label{fig:deform_potts}
\end{center}
\end{figure*}

\begin{figure*}[h]
\begin{center}
\includegraphics[width=\textwidth]{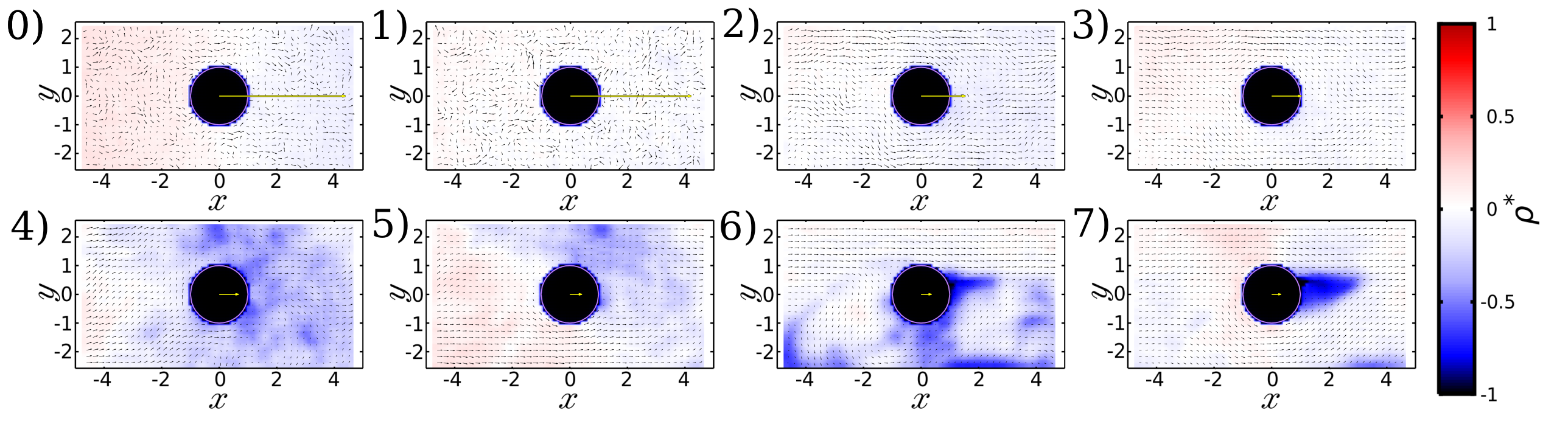}
\caption{Multiparticle model: density and velocity, for the 8 cases outlined in Tables~\ref{tab:param-super} and \ref{tab:multiparticle-cube-result}.}
\label{fig:den_eq_vel_super}
\end{center}
\end{figure*}

\begin{figure*}[h]
\begin{center}
\includegraphics[width=\textwidth]{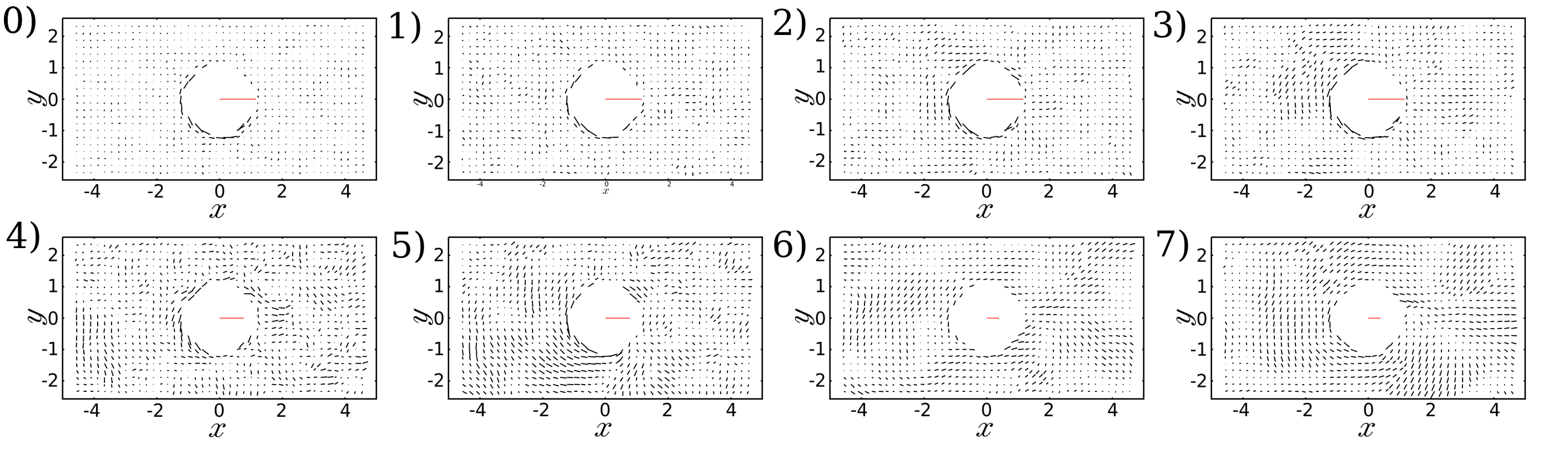}
\caption{Multiparticle model: deformation anisotropy, for the 8 cases outlined in Tables~\ref{tab:param-super} and \ref{tab:multiparticle-cube-result}.}
\label{fig:deform_super}
\end{center}
\end{figure*}

\section{Results}
\label{maps}

Figs.~\ref{fig:den_eq_vel_vicsek}-\ref{fig:deform_super} 
represent the output measurement maps for the five models, using the scheme explained above:
In Figs. \ref{fig:den_eq_vel_vicsek}, \ref{fig:den_eq_vel_szabo}, \ref{fig:den_eq_vel_voronoi}, \ref{fig:den_eq_vel_potts}, and \ref{fig:den_eq_vel_super},  the normalized density $\delta \rho$ is in color, with blue, white and red representing density lower, equal and higher than the equilibrium density $\rho_{\text{eq}}$, respectively.
Velocity is represented as black arrows on the same plot, with a yellow unit scale arrow shown in the middle of the obstacle. When the flow is slow and disordered the scale appears large, while when a strongly collective flow is established the scale appears small.
In Figs. \ref{fig:deform_vel_vicsek}, \ref{fig:deform_vel_szabo}, \ref{fig:deform_voronoi}, \ref{fig:deform_potts} and \ref{fig:deform_super}, the deformation anisotropy magnitude and direction are represented by a bar length and direction. To indicate the scale, the red line in the middle of the obstacle represents the deformation $\ln 2$, corresponding to cells whose length is twice their width. This means when the cell deformations are small and disordered, the scale appears large, while when a strongly collective deformation pattern is established the scale appears small.



\LB{ Figure~\ref{fig:den_eq_vel_vicsek} presents velocity-density maps for the Vicsek model. At low alignment, the density is higher before the obstacle, and its distribution is symmetrical around the $y$-axis. In contrast, at high alignment and low densities (Figs.~\ref{fig:den_eq_vel_vicsek}-4 and ~\ref{fig:den_eq_vel_vicsek}-6), there is a break in the density distribution symmetry, and the region of high density shifts towards one of the narrow spaces between the obstacle and the wall. The deformation maps (\ref{fig:deform_vel_vicsek}) indicate an upstream/downstream asymmetry, particularly near the obstacle, where the direction close to it is favored. When compared to the other cases, the low alignment and low attraction cases (maps 0 and 1) exhibit minimal deformations.}


\LB{Figures~\ref{fig:den_eq_vel_szabo} and \ref{fig:deform_vel_szabo} illustrate the Szabo model's behavior. In all cases, the density map appears to be roughly symmetrical with respect to the $y$-axis, with high densities before the obstacle. The deformations observed in this model are more intense than those in the Vicsek model and display different patterns before the obstacle. When the alignment is low (maps 0-4), the deformation is primarily in the $x$-direction. In contrast, at high alignment, except for the high-density and low adhesion case (map  5), there is a region of low deformation just downstream of the obstacle. However, upstream of the obstacle, a region of high deformation can be observed.
}


\LB{The Voronoi model, as depicted in Figs.~\ref{fig:den_eq_vel_voronoi} and \ref{fig:deform_voronoi}, exhibits significant density fluctuations with explicit alignment. Similar to the Vicsek model, low alignment results in small deformations, which increase significantly with both types of alignments. Additionally, deformations are closely tangential to the obstacle in most cases.
}


\LB{Figures~\ref{fig:den_eq_vel_potts} and \ref{fig:deform_potts} demonstrate the Potts model's behavior. With the exception of map 0, the density is higher before the obstacle, and some accumulation occurs close to the obstacle in case of low adhesion and high alignment (map 4). Furthermore, a clear contrast in the velocity field is evident in cases with and without alignment (maps 0,1,2,3 versus maps 4,5,6,7). The deformations are primarily tangential to the obstacle in the upstream and lateral parts of the obstacle. However, they are parameter-dependent in the downstream part, as demonstrated by the comparison between maps 4 and 6 on one side and 3 and 5 on the other.
}


\LB{Within the parameter range investigated in this study, the Multiparticle model (Figs.~\ref{fig:den_eq_vel_super}, \ref{fig:deform_super}) exhibits coherent polarization waves but does not exhibit complete ordering. The density exhibits limited variation, and fluctuations arise due to averaging over void regions. Tangential deformation is dominant near the obstacle, although voids can alter this behavior.
}

\section{Discussion: Choosing a model}

For each given scientific question, several criteria can help to choose a suitable numerical model. In order to help the reader, we provide several comparison tables.   
Table~\ref{tab:qualitative} provides an overview of the physical ingredients incorporated in each model. 

\begin{table*}[h]
    \centering
    \begin{tabular}{|c|c|c|c|c|c|}
            \hline 
            & Vicsek & Szab\'o & Voronoi & Potts & Multiparticle  \\
            \hline
            \hline
            Degree of freedom &  Particle &  Particle & Particle & Contours & Multiparticle  \\
            Cell shape & Disk & Disk & Polygon & Set of pixels & Polygon \\
            Alignment & Neighbor & Persistence & Both & Both & Both  \\
            Interaction & Force & Force & Tension & Tension & Force  \\
            Core & Hard  & Soft  & Soft  & None & Hard \\
            Lattice & No & No & No & Yes & No  \\
            Speed & Fixed & Variable & Variable & Variable & Fixed  \\
            Walls & Repulsive & Repulsive & Attractive & Attractive & Repulsive  \\
            Obstacle & Repulsive & Friction & Attractive & Attractive  & Repulsive  \\
            Cell source  & Creation & Creation & Division  & Division & Division \\
            Cell death & Yes & Yes & No   & Yes & Yes  \\
            \hline
    \end{tabular}
\caption{Overview of model ingredients. ``Force" refers to pairwise radial forces between cell centers, while ``tension" refers to cell-cell junction tension. }
    \label{tab:qualitative}
\end{table*}

Table~\ref{tab:parameters_limitations}  explains how to choose the model parameter in order to avoid artifacts and execution troubles.
 For instance, some of the parameters only make sense for 
    positive values, such as the alignment or the 
    force between cells.
 In some cases, if the interaction or the junction tension is too strong, the cells can shrink or even disappear. In the Potts and Multiparticle models, cells may break if the alignment parameter is excessive.

 In all models the density should be carefully adjusted:
In the Voronoi model, which is always confluent by construction (no free space is allowed), a too low density induces  very unrealistic  cell shapes and velocities.
In the other models a too low density prevents confluence, i.e. some cells form small groups surrounded by free space. Conversely, at high density in all models except for Potts, the pressure becomes too high and induces spurious movements, cell overlaps or obstacle invasion. In particular, in the Multiparticle model cells easily overlap which induces severe artifacts.

\begin{table*}
    
    \centering
    \begin{tabular}{|c|c|c|c|c|c|}
            \hline 
            & Vicsek & Szab\'o & Voronoi & Potts & Multiparticle  \\
            \hline
            \hline
            Lowest & $0.0$ & N/A & $0.0$ & $0.0$ & $0.0$ \\
            alignment & {\it disorder} &   &  {\it disorder} &  {\it disorder} &  {\it disorder} \\
            \hline
          Highest & no limit & N/A & finite & finite &  finite \\
            alignment &   &  & {\it alignment drops} & {\it cells break} & {\it cells break} \\
            \hline
            Lowest & N/A & $1/\tau \to 0$ & $1/\tau \to 0$ &  0.0 &  0.0 \\
            persistence &  &  {\it disorder} &  {\it disorder} &  {\it disorder} &  {\it disorder} \\
            \hline
            Highest & N/A & finite & finite & finite & finite  \\
            persistence &  & {\it alignment drops} & {\it tissue collapse} & {\it cells break} & {\it cells break} \\
            \hline
            Lowest & finite & finite & finite & finite & finite \\
            interaction & {\it gas} & {\it gas} & {\it rosettes} & {\it cells break} & {\it gas} \\
            \hline
            Highest & core repulsion & repulsion force & finite & finite & finite \\
            interaction &   {\it cores overlap} &  {\it cores overlap} &  {\it cells collapse}  &  {\it cells collapse} & {\it cells break} \\
            \hline
            Lowest & confluence & confluence & confluence & confluence & confluence \\
          density & {\it holes} &  {\it holes} & {\it retraction} &  {\it holes} &  {\it holes} \\
            \hline
            Highest & finite & finite & finite & finite & 
            finite \\
            density &  {\it cores overlap,} &  {\it cores overlap} & {\it cells are liquid,} & {\it cells collapse} &  {\it cells collapse} \\
            & {\it invade obstacle} &  & {\it execution crashes} & & {\it and/or overlap} \\
            \hline
    \end{tabular}
    \caption{Model parameter limitations. For each model, lower and upper limits are suggested in the top line. For each limit,  a reason for this choice (e.g. the appearance of an artifact) is indicated in {\it italics} in the bottom line. Here we use N/A: not applicable.
}
    \label{tab:parameters_limitations}
\end{table*}

Table~\ref{tab:model_measure} compares the range of input measurements range that each model can reasonably simulate. Since these input measurements are standardized and dimensionless, this comparison is physically relevant. For instance, all models enable us to vary alignment $\phi$, but the Vicsek model can produce high alignment, while the Potts  and Multiparticle models are restricted to smaller values of $\phi$ to remain stable. 
The Multiparticle model is suitable for low density simulations; in fact, density falls bellow the equilibrium one when in collective motion.
In the other four models it is possible to increase the density above the equilibrium value by controlling the cell creation rate. The Voronoi model is the most suitable for reaching a high density. 
     
All models reasonably reproduce both liquid and solid behaviours, although this can be sensitive to alignment, to forces and to several artifacts. More precisely, in the absence of collective behavior (low $\phi$), Vicsek, Szab\'o and Potts models 
present $\Delta$ values which increase with the force. In collective motion (high $\phi$), the Vicsek, Szab\'o and Potts models show solid behaviour (high $\Delta$) independently of attraction forces or density. Conversely, the Voronoi model displays a liquid behavior (low $\Delta$) at high densities, whatever the force; while in the Multiparticle model, at high force cell shapes become more irregular, neighbor exchanges become more frequent and thus the behaviour becomes liquid ($\Delta$ decreases).

\begin{table*}[h]
    \centering
    \begin{tabular}{|c|c|c|c|c|c|}
            \hline 
            & Vicsek & Szab\'o & Voronoi & Potts & Multiparticle  \\
            \hline
            \hline
            Alignment &  &  &  &  &  \\
            low $\phi$           & $0.12$ & $0.2$ & $0.17$ & $0.09$ & $0.2$ \\ 
            high $\phi$          & $0.98$ & $0.9$ & $0.9$  & $0.5$  & $0.6$ \\  
            \hline
            Density &  &  &  &  &  \\
            low $\delta\rho$    & $0.05$ & $0.2$ & $0.5$   & $0.25$ & $-0.1$ \\
            high $\delta\rho$   & $1.0$  & $1.0$ & $1.1$   & $1.1$  & $0.1$ \\
            \hline
            Liquid / solid &  &  &  &  &  \\
            low $\Delta$       & $0.02$ & $0.2$  & $0.2$   & $0.09$ & $0.1$ \\
            high $\Delta$      & $0.85$ & $0.85$ & $0.7$   & $0.8$  & $0.7$ \\
            \hline
    \end{tabular}
    \caption{Input measurements range reached for each model. 
    All values indicated are approximate. Remember that the normalized density is compared with the equilibrium density (Eq.~\ref{rho}), hence can reach negative values. }
    \label{tab:model_measure}
\end{table*}

Finally, as expected, the Vicsek and Szab\'o models are simple and robust. Conversely, the Potts and Multiparticle models offer realistic shapes, shape changes and neighbour exchanges. In between, the Voronoi model is often a good compromise. Table~\ref{tab:model_properties_comparison} refines this comparison. These appreciations are entirely subjective and solely intended to help in choosing a suitable model. Criteria include the physical ingredients, parameter limitations, quantities to be measured, possible artifacts, simulation running time, and even the likelihood of execution crashes. For instance, depending on the flow alignment and spatial gradients, after the obstacle a hole can appear (or, in the Voronoi model, cell shapes and velocities become unrealistic).

\begin{table*}[h]
    \centering
    \begin{tabular}{|c|c|c|c|c|c|}
            \hline 
           & Vicsek & Szab\'o & Voronoi & Potts & Multiparticle  \\
           \hline
           \hline
            Density range &
           {\protect\footnotesize  \textcolor{yellow!40!orange}{\faStar\faStar\faStar\faStar\faStar}} &
           {\footnotesize  \textcolor{yellow!40!orange}{\faStar\faStar\faStar\faStar\faStar}} &
           {\footnotesize  \textcolor{yellow!40!orange}{\faStar\faStar}\textcolor{white!70!black}{\faStarO\faStarO\faStarO}} & 
           {\footnotesize  \textcolor{yellow!40!orange}{\faStar\faStar\faStar}\textcolor{white!70!black}{\faStarO\faStarO}} &
           {\footnotesize  \textcolor{yellow!40!orange}{\faStarHalfO}\textcolor{white!70!black}{\faStarO\faStarO\faStarO\faStarO}} \\
            \hline 
            Density fluctuations  & 
           {\footnotesize  \textcolor{yellow!40!orange}{\faStar\faStar\faStar\faStar\faStar}} &
           {\footnotesize  \textcolor{yellow!40!orange}{\faStar\faStar}\textcolor{white!70!black}{\faStarO\faStarO\faStarO}} & 
           {\footnotesize  \textcolor{yellow!40!orange}{\faStar\faStar\faStar}\textcolor{white!70!black}{\faStarO\faStarO}} & 
           {\footnotesize  \textcolor{yellow!40!orange}{\faStar\faStar\faStar}\textcolor{white!70!black}{\faStarO\faStarO}} &
           {\footnotesize  \textcolor{yellow!40!orange}{\faStar}\textcolor{white!70!black}{\faStarO\faStarO\faStarO\faStarO}} \\
            \hline 
            Tissue shrinkage & 
           {\footnotesize  \textcolor{yellow!40!orange}{\faStarHalfO}\textcolor{white!70!black}{\faStarO\faStarO\faStarO\faStarO}} & 
           {\footnotesize  \textcolor{yellow!40!orange}{\faStar}\textcolor{white!70!black}{\faStarO\faStarO\faStarO\faStarO}} & 
           {\footnotesize  \textcolor{yellow!40!orange}{\faStar\faStar\faStar\faStar\faStar}} & 
           {\footnotesize  \textcolor{yellow!40!orange}{\faStar\faStar\faStar\faStar}\textcolor{white!70!black}{\faStarO}} & 
           {\footnotesize  \textcolor{yellow!40!orange}{\faStar\faStar}\textcolor{white!70!black}{\faStarO\faStarO\faStarO}} \\
            \hline
            Cell shape & 
           {\footnotesize  \textcolor{yellow!40!orange}{\faStar}\textcolor{white!70!black}{\faStarO\faStarO\faStarO\faStarO}} &
           {\footnotesize  \textcolor{yellow!40!orange}{\faStar}\textcolor{white!70!black}{\faStarO\faStarO\faStarO\faStarO}} &
           {\footnotesize  \textcolor{yellow!40!orange}{\faStar\faStar\faStar}\textcolor{white!70!black}{\faStarO\faStarO}} &
           {\footnotesize  \textcolor{yellow!40!orange}{\faStar\faStar\faStar\faStar\faStar}} &
           {\footnotesize  \textcolor{yellow!40!orange}{\faStar\faStar\faStar\faStar\faStar}} \\
            \hline
            Cell stretch & 
           {\footnotesize  \textcolor{yellow!40!orange}{\faStarHalfO}\textcolor{white!70!black}{\faStarO\faStarO\faStarO\faStarO}} & 
           {\footnotesize  \textcolor{yellow!40!orange}{\faStarHalfO}\textcolor{white!70!black}{\faStarO\faStarO\faStarO\faStarO}} & 
           {\footnotesize  \textcolor{yellow!40!orange}{\faStar\faStar\faStar\faStar\faStar}} & 
           {\footnotesize  \textcolor{yellow!40!orange}{\faStar\faStar\faStarHalfO}\textcolor{white!70!black}{\faStarO\faStarO}} & 
           {\footnotesize  \textcolor{yellow!40!orange}{\faStar\faStar\faStar}\textcolor{white!70!black}{\faStarO\faStarO}} \\
            \hline
            Cell velocity & 
           {\footnotesize  \textcolor{yellow!40!orange}{\faStar\faStarHalfO}\textcolor{white!70!black}{\faStarO\faStarO\faStarO}} &
           {\footnotesize  \textcolor{yellow!40!orange}{\faStar\faStar}\textcolor{white!70!black}{\faStarO\faStarO\faStarO}} &
           {\footnotesize  \textcolor{yellow!40!orange}{\faStar\faStar\faStar}\textcolor{white!70!black}{\faStarO\faStarO}} &
           {\footnotesize  \textcolor{yellow!40!orange}{\faStar\faStar}\textcolor{white!70!black}{\faStarO\faStarO\faStarO}} &
           {\footnotesize  \textcolor{yellow!40!orange}{\faStar\faStarHalfO}\textcolor{white!70!black}{\faStarO\faStarO\faStarO}} \\
            \hline
            Velocity asymmetry & 
           {\footnotesize  \textcolor{yellow!40!orange}{\faStar\faStarHalfO}\textcolor{white!70!black}{\faStarO\faStarO\faStarO}} &
           {\footnotesize  \textcolor{yellow!40!orange}{\faStar\faStar}\textcolor{white!70!black}{\faStarO\faStarO\faStarO}} &
           {\footnotesize  \textcolor{yellow!40!orange}{\faStar\faStar\faStar\faStar\faStar}} &
           {\footnotesize  \textcolor{yellow!40!orange}{\faStar\faStar}\textcolor{white!70!black}{\faStarO\faStarO\faStarO}} &
           {\footnotesize  \textcolor{yellow!40!orange}{\faStar\faStarHalfO}\textcolor{white!70!black}{\faStarO\faStarO\faStarO}} \\
            \hline
            Cell self-persistence & 
           {\footnotesize  \textcolor{white!70!black}{\faStarO\faStarO\faStarO\faStarO\faStarO}} &
           {\footnotesize  \textcolor{yellow!40!orange}{\faStar\faStar\faStar\faStar}\textcolor{white!70!black}{\faStarO}} &
           {\footnotesize  \textcolor{yellow!40!orange}{\faStar\faStar\faStar\faStar\faStar}} &
           {\footnotesize  \textcolor{yellow!40!orange}{\faStar\faStar\faStar}\textcolor{white!70!black}{\faStarO\faStarO}} &
           {\footnotesize  \textcolor{yellow!40!orange}{\faStar\faStar\faStar}\textcolor{white!70!black}{\faStarO\faStarO}} \\
            \hline
            Neighbour alignment & 
           {\footnotesize  \textcolor{yellow!40!orange}{\faStar\faStar\faStar\faStar\faStar}} &
           {\footnotesize  \textcolor{yellow!40!orange}{\faStar\faStar\faStar\faStar}\textcolor{white!70!black}{\faStarO}} &
           {\footnotesize  \textcolor{yellow!40!orange}{\faStar\faStar\faStar\faStarHalfO}\textcolor{white!70!black}{\faStarO}} &
           {\footnotesize  \textcolor{yellow!40!orange}{\faStar\faStar}\textcolor{white!70!black}{\faStarO\faStarO\faStarO}} &
           {\footnotesize  \textcolor{yellow!40!orange}{\faStar\faStar\faStar\faStar}\textcolor{white!70!black}{\faStarO}} \\
            \hline
            Neighbour exchange & 
           {\footnotesize  \textcolor{yellow!40!orange}{\faStar\faStar}\textcolor{white!70!black}{\faStarO\faStarO\faStarO}} &
           {\footnotesize  \textcolor{yellow!40!orange}{\faStar\faStar}\textcolor{white!70!black}{\faStarO\faStarO\faStarO}} &
           {\footnotesize  \textcolor{yellow!40!orange}{\faStar\faStar\faStar}\textcolor{white!70!black}{\faStarO\faStarO}} &
           {\footnotesize  \textcolor{yellow!40!orange}{\faStar\faStar\faStar\faStar\faStar}} &
           {\footnotesize  \textcolor{yellow!40!orange}{\faStar\faStar\faStar\faStar}\textcolor{white!70!black}{\faStarO}} \\
            \hline
           Closing after obstacle & 
           {\footnotesize  \textcolor{yellow!40!orange}{\faStar}\textcolor{white!70!black}{\faStarO\faStarO\faStarO\faStarO}} &
           {\footnotesize  \textcolor{yellow!40!orange}{\faStar\faStar\faStarHalfO}\textcolor{white!70!black}{\faStarO\faStarO}} &
           {\footnotesize  \textcolor{yellow!40!orange}{\faStar\faStar}\textcolor{white!70!black}{\faStarO\faStarO\faStarO}} &
           {\footnotesize  \textcolor{yellow!40!orange}{\faStar\faStar\faStar\faStar\faStar}} &
           {\footnotesize  \textcolor{yellow!40!orange}{\faStar\faStar\faStar\faStar}\textcolor{white!70!black}{\faStarO}} \\
            \hline
            Simulation time efficiency & 
           {\footnotesize  \textcolor{yellow!40!orange}{\faStar\faStar\faStar\faStar\faStar}} &
           {\footnotesize  \textcolor{yellow!40!orange}{\faStar\faStar\faStar\faStar\faStar}} &
           {\footnotesize  \textcolor{yellow!40!orange}{\faStar\faStar\faStarHalfO}\textcolor{white!70!black}{\faStarO\faStarO}} &
           {\footnotesize  \textcolor{yellow!40!orange}{\faStarHalfO}\textcolor{white!70!black}{\faStarO\faStarO\faStarO\faStarO}} &
           {\footnotesize  \textcolor{yellow!40!orange}{\faStar\faStarHalfO}\textcolor{white!70!black}{\faStarO\faStarO\faStarO}} \\
            \hline
            Simulation stability & 
           {\footnotesize  \textcolor{yellow!40!orange}{\faStar\faStar\faStar\faStar}\textcolor{white!70!black}{\faStarO}} &
           {\footnotesize  \textcolor{yellow!40!orange}{\faStar\faStar\faStar\faStar}\textcolor{white!70!black}{\faStarO}} &
           {\footnotesize  \textcolor{yellow!40!orange}{\faStar\faStar}\textcolor{white!70!black}{\faStarO\faStarO\faStarO}} &
           {\footnotesize  \textcolor{yellow!40!orange}{\faStar\faStar\faStar}\textcolor{white!70!black}{\faStarO\faStarO}} &
           {\footnotesize  \textcolor{yellow!40!orange}{\faStar\faStar\faStar}\textcolor{white!70!black}{\faStarO\faStarO}} \\
            \hline
            \hline
            Number of cells   & $\sim 7600 $ &   $\sim 11400 $     & $\sim 8100$  & $\sim 7500$ & $\sim 7300$ \\ 
            \hline
            Main  advantages & robustness, & simplicity, & good & shape, & large \\
             & simplicity & alignment & compromise  &  fluctuations & deformations \\
            \hline
            Main limitation & shape & shape & density & alignment & density \\
           \hline
    \end{tabular}
    \caption{Model guide chart: subjective appreciations of each model's advantages (5 colored stars indicate the best quality). 
    }
    \label{tab:model_properties_comparison}

\end{table*}




%
%
%



\section{Conflicts of Interest}
There are no conflicts of interest to declare.

\section{Acknowledgments}

This project has been funded by CAPES-COFECUB Ph 880-17 "From cell to tissue: collective mechanical behaviours". 
C.B. has been funded by CAPES and by ANR "Migrafolds".
We thank \FG{M. Durand for critical reading of the manuscript. We thank S. Tlili for Fig. 1; and} R. de Almeida, G. Thomas, M. Durande, S. Tlili, H. Delano\"e-Ayari for discussions. We thank R. Sknepnek for his guidance in setting up the Voronoi simulation using SAMoS~\cite{SAMoS}. We dedicate this work to the memory of C. Kirch.

\clearpage

\nolinenumbers

%
%
%
\bibliography{stokes-bench}
\bibliographystyle{stokes-bench}

\end{document}